\documentclass[reprint,amsmath,amssymb,textcomp,aps]{revtex4-1}
\usepackage{amsmath}
\usepackage{amssymb}
\usepackage{textcomp}
\usepackage{graphicx}
\usepackage{dcolumn}
\usepackage{bm}

\begin{document}

\preprint{APS/123-QED}

\title{Trion gas, electron-hole liquid, and metal--insulator transition in doped heterostructures based on transition metal dichalcogenides}

\author{P.~V.~Ratnikov}
\email{ratnikov@lpi.ru}
\affiliation{A.~M.~Prokhorov General Physics Institute, Russian Academy of Sciences, ul. Vavilova 38, 119991 Moscow, Russia}

\author{\fbox{A.~P.~Silin}}
%\email{a.p.silin@mail.ru}
\affiliation{P.~N.~Lebedev Physical Institute, Russian Academy of Sciences, Leninskii pr. 53, 119991 Moscow, Russia\\
Moscow Institute of Physics and Technology (National Research University), Dolgoprudnyi, 141700 Moscow region, Russia}

\date{\today}

\begin{abstract}
The effect of doping on the parameters of an electron-hole liquid (EHL) in heterostructures based on transition metal dichalcogenides is studied. The phase diagram of the EHL is constructed. It is shown that for the formation of a high-temperature tightly bound EHL, as well as for the transition from the semiconducting (exciton) state to the semimetallic one (electron-hole plasma/liquid), it is advantageous to dope the energy band with larger number of valleys. The transition from trion gas to electron-hole plasma is investigated using the modified Mott criterion and variational calculation with screened potential. The effect of doping on the metal--insulator transition in the equilibrium case without laser excitation is studied.
\end{abstract}

\pacs{71.35.−y, 73.21.Fg, 73.90.+f}

\keywords{2D materials, heterostructures, dichalcogenides, electron-hole liquid, excitons, trions}
\maketitle

\section{\label{s1}Introduction}
The beginning of the 21st century was characterized by the rapid growth of nanotechnology using two-dimensional (2D) materials such as graphene and transition metal dichalcogenides (TMDs). Graphene nanoelectronics has been actively developed \cite{RS2018}. Graphene is now often used as flexible, durable and transparent electrodes. For example, graphene electrodes are used in the manufacture of solar cells based on perovskites \cite{Wang2014}. Unfortunately, graphene-based nanoelectronics has limited applicability. On the contrary, TMDs are now considered extremely useful for the further development of logical devices \cite{Ahmed2017}.

TMDs have the general chemical formula $MX_2$ with a transition metal atom $M$ (Nb, Ta, Mo, W, Re, etc.) and two chalcogen atoms $X$ (S, Se, Te). Bulk samples are a stack of monomolecular layers (monolayers) connected by van der Waals forces, similar to graphite viewed as a stack of graphene layers. The monolayers are separated by $\approx$6.5 \AA. Each monolayer has the form of a sandwich with a layer of $M$ atoms inserted between two layers of $X$ atoms. The most common crystal structures of TMDs are trigonal prismatic (2H) and octahedral (1T) \cite{Dickinson1923}.

The most TMDs, like MoS$_2$ and especially ReS$_2$ and ReSe$_2$, are characterized by weak interaction between the layers. They demonstrate a small change (about twofold increase) in the band gap as they change from three-dimensional (3D) samples to monolayer films. On the contrary, the TMDs of the tenth group exhibit a strong interaction between the layers. For example, PtS$_2$ has the band gap 0.25 eV in a bulk and 1.6 eV in a monolayer and a clearly pronounced dependence of the band gap on the number of layers is observed like black phosphorus~\cite{Tran2014}.

A detailed description of the structure and synthesis of TMDs is presented in the review \cite{Chern2018}.

Low-temperature doped electron-hole liquid (EHL) in traditional semiconductors (Ge and Si) has been experimentally and theoretically studied (see the monograph \cite{JeffrisKeldysh1988} and references therein). The density of excess electrons or holes was determined by the number of donor or acceptor atoms, thence it could not be controlled in the considered sample. It changed only when the sample was replaced.

The situation changed significantly with using monolayer or bilayer heterostructures based on TMDs. In addition to the fact that EHL becomes high-temperature and tightly bound \cite{Yu2019, Pekh2020}, the density of excess charge carriers is determined not by the number of introduced impurity atoms, but by electron or hole doping, when the carrier density is changed with the help of an electric field effect. A dielectric substrate (usually heavily doped silicon with a SiO$_2$ layer) acts as a gate in a field-effect transistor \cite{Novoselov2004}. The electric field effect causes electrons to flow into the 2D material when a positive voltage is applied to the gate. This is an electron doping of the 2D material. With a negative gate voltage, an excess of holes appears in the 2D material (a hole doping occurs).

There are alternative ionic approaches to controlling the carrier density in 2D materials.

The ionic gating technique is based on the suction of mobile charges from ion gel with the formation of electric double layers. With the help of ionic gating, ultra-high carrier densities (over $10^{14}$ cm$^{- 2}$) are attainable. The usage of 2D TMDs in this technique is especially convenient due to their atomically smooth surfaces \cite{Liu2017}.

A doping method using the gate-controlled Li ion intercalation into thin films 1T-TaS$_2$ was developed for creation of ionic field-effect transistor \cite{Yu2015}.

When Rb ions are deposited on the surface of 3D MoSe$_2$ samples, a built-in electric field appears, which shifts the Fermi level up to the conduction band. There is a limit $1.4\times10^{14}$ cm$^{-2}$ in this method on the carrier density. In a stronger electric field, corresponding to the density $1.5\times10^{14}$ cm$^{-14}$, an~indirect to direct band gap transition is induced \cite{Kim2017}.

Calculations using the density functional theory showed that the band gaps in the MoS$_2$, MoSe$_2$, MoTe$_2$, and WS$_2$ bilayers are continuously decreasing when the built-in electric field increases \cite{Rama2011}. For MoS$_2$ in an electric field of 3 V/nm, the conduction and valence bands intersect, and at an electric field about 2 V/nm, the MoS$_2$ bilayer becomes direct-gap semiconductor (in zero field, it is an indirect-gap one).

Such a restructuring of the band gap was experimentally implemented in a field-effect transistor based on a MoS$_2$ bilayer \cite{Chu2015}. The decrease in the band gap occurred linearly by $\approx$260 meV with an increase in the electric field by 1 V/nm.

When charge carriers are added via the electric field effect, excitons ($X^0$) in monolayers MoS$_2$ \cite{Mak2013}, MoSe$_2$ \cite{Ross2013}, and MoTe$_2$ \cite{Yang2015} are transformed into trions, electron-hole ($e$-$h$) triples, consisting of two electrons and one hole (electron trion $X^-$) or two holes and one electron (hole trion $X^+$). At the same time, the dependence of the absorption and photoluminescence spectra on the carrier density was investigated from the neutrality point (without doping) up to a doping density of $\sim10^{13}$ cm$^{-2}$. The exciton peak decreased and disappeared, merging with the background, while the trion lines remained, gradually broadening and approximately preserving its spectral weight.

A decrease in the exciton binding energy within 100 meV was observed in WS$_2$. It associated with screening of the Coulomb potential by injected free charge carriers with a density up to $8\times10^{12}$ cm$^{-2}$. Estimates show that excitons are completely ionized at a carrier density of $\sim10^{13}$ cm$^{-2}$ \cite{Chernikov2015a}.

In previous works \cite{Pekh2020, Pekh2021}, we proved that the multivalley structure of the energy spectrum in TMD monolayers leads to a noticeable increase in the equilibrium density and binding energy of EHL.

EHL and the metal--insulator transition in doped multivalley semiconductors were considered in the framework of the high anisotropy model \cite{Andryushin1976a} in the works \cite{Andryushin1990a, Andryushin1990b, Andryushin1991}. Correlation effects play an anomalously large role. It~was shown that the equilibrium EHL density decreases, the binding energy and equilibrium chemical potential increase in absolute value with increasing doping. EHL does not arise at sufficiently strong doping.

Doping of EHL significantly complicates the density--temperature phase diagram for the gas--liquid and metal--insulator transitions. In this case, in addition to temperature, there are two densities, the density of $e$-$h$ pairs $n$ determined by optical excitation (the light absorbed in the TMD layer) and the density of excess electrons or holes $\Delta n$ controlled by the static voltage applied to the heterostructure gate.

Note the interesting case $\Delta n=n$. Then, unlike the usual undoped $e$-$h$ system, consisting of $e$-$h$ pairs or excitons, doped $e$-$h$ system consists of $e$-$h$ triples (trions). Trions have a finite, but relatively low (compared to an exciton) binding energy. The binding energy of an exciton in TMDs is hundreds of meV, and a trion energy is tens of meV (in bulk semiconductors, the binding energy of trions is very small) \cite{Durnev2018}. Therefore, the gas branch (with a lower density) of the phase diagram corresponds to a trion gas, and a liquid branch (with a higher density) corresponds to a doped EHL.

The gas is not purely trionic, if $\Delta n\neq n$. We consider electron doping for definiteness. At $\Delta n<n$, there are not enough electrons and the gas is a mixture of $X^0$ and $X^-$. At $\Delta n>n$, electrons are in excess and the gas is a mixture of $X^-$ and electrons. The screening of trions by electrons can lead to the disappearance of the bound state.

In this work, we investigate the effect of doping on the parameters of EHL and characteristics of the metal--insulator transition in heterostructures based on TMDs.

\section{\label{s2}Ground state energy of doped EHL}

We consider a model 2D semiconductor with $\nu_e$ electron and $\nu_h$ hole valleys. The electron and hole densities are $n_e$ and $n_h$, respectively. Let be $n_e\geq n_h$. The excess of the electron density over the hole density $\Delta n =n_e-n_h$ is set by the gate voltage. Under the conditions of photoexcitation, the variable quantity is the density of generated $e$-$h$ pairs $n=n_h$. For a given intensity of photoexcitation, it is convenient to introduce the parameter $\delta=\Delta n/n$ (electron doping level).

The carrier density is measured in $a^{-2}_x$, where $a_x$ is the Bohr radius of 2D exciton,
\begin{equation*}
a_x=\frac{\hbar^2}{2m\widetilde{e}^2}.
\end{equation*}
The energy is measured in units of the 2D exciton binding energy
\begin{equation*}
E_x=\frac{2m\widetilde{e}^4}{\hbar^2}.
\end{equation*}
Here, $m=m_em_h/(m_e+m_h)$ is the reduced mass of an electron $m_e$ and a hole $m_h$, $\widetilde{e}^2=e^2/\varepsilon_\text{eff}$ and $\varepsilon_\text{eff}=(\varepsilon_1+\varepsilon_2)/2$ is the effective permittivity, $\varepsilon_1$ and $\varepsilon_2$ are low-frequency permittivities of the surrounding media. In what follows, Planck's constant $\hbar$ will be assumed to be equal to unity.

We introduce the dimensionless distance between particles (in this case, holes)
\begin{equation*}
r_s=\sqrt{\frac{\nu_h}{\pi n}}.
\end{equation*}

Wave vectors are measured in units of the Fermi wave vector of holes
\begin{equation*}
q^h_F=\sqrt{\frac{2\pi n}{\nu_h}}=\frac{\sqrt{2}}{r_s}.
\end{equation*}

We represent the ground state energy in the form ($\varkappa=\nu_e/\nu_h$ and $\sigma=m_e/m_h$)
\begin{equation}\label{E_gs}
\begin{split}
E_\text{gs}=&\frac{1}{(1+\sigma)r^2_s}\left(\sigma+\frac{1+\delta}{\varkappa}\right)\\
&-\frac{4\sqrt{2}}{3\pi r_s}\left(1+\sqrt{\frac{1+\delta}{\varkappa}}\right)+\int\limits_0^\infty I(q)dq.
\end{split}
\end{equation}
The first two terms are the energy in the Hartree--Fock approximation (kinetic and exchange energy). The last term is a correlation energy written as an integral over the wave vector $q$. It is divided by the average particle density $\overline{n}=(n_e+n_h)/2=\left(1+\frac{\delta}{2}\right)n$.

The function $I(q)$ for $q\ll1$ is determined in the random phase approximation (RPA), and it for $q\gg1$ is the sum of second-order diagrams in the interaction (loop and exchange ones):
\begin{equation}\label{Iq}
I(q)=\begin{cases}
-aq+bq^{3/2}-cq^2+dq^{5/2}+fq^3, & q\ll1,\\
-gq^{-3},& q\gg1.
\end{cases}
\end{equation}

For most TMDs, it turned out to be $m_e\approx m_h$. It~is~well also known that the dependence of the EHL parameters on the mass ratio is weak \cite{Pekh2021}. Therefore, we present expressions for the coefficients in the case of equal masses of the electron and hole ($\sigma=1$)
\begin{equation*}
\begin{split}
a&=\frac{\sqrt{2}}{\left(1+\frac{\delta}{2}\right)\pi r_s}\left[1+\sqrt{\varkappa(1+\delta)}\right],\\
b&=\frac{2^{1/4}}{\left(1+\frac{\delta}{2}\right)^{1/2}r^{3/2}_s\nu^{1/2}_h},\\
c&=\frac{\pi\Theta(\varkappa,\,\delta)-f_1(\varkappa,\,\delta)-f_2(\varkappa,\,\delta)}{\left(1+\frac{\delta}{2}\right)\pi r^2_s\nu_h},\\
d&=\frac{3}{2^{17/4}\left(1+\frac{\delta}{2}\right)^{5/2}r^{5/2}_s\nu^{3/2}_h}\left(1+\frac{(1+\delta)^2}{\varkappa}\right),
\end{split}
\end{equation*}
\begin{widetext}
\begin{equation*}
\begin{split}
f&=\frac{1}{\sqrt{2}\left(1+\frac{\delta}{2}\right)\pi r^3_s}\left\{\frac{r^2_s}{12}\left(1+\frac{\varkappa^{3/2}}{\sqrt{1+\delta}}\right)-
\frac{1}{\nu^2_h}\left[\frac{f_3(\varkappa,\,\delta)}{2(1+\varkappa)^2}+f_4(\varkappa,\,\delta)\right]\right\},\\
g&=\frac{1}{1+\frac{\delta}{2}}\left[2\nu_h-1+4\nu_h(1+\delta)+\left(2\nu_h-\frac{1}{\varkappa}\right)(1+\delta)^2\right],
\end{split}
\end{equation*}
where the functions are introduced
\begin{equation*}
\begin{split}
f_1(\varkappa,\,\delta)&=\int\limits_0^{\min\{\sqrt{\frac{1+\delta}{\varkappa}},\,1\}}\arctan\left[\frac{x}{1+\varkappa}
\left(\frac{\varkappa}{\sqrt{\frac{1+\delta}{\varkappa}-x^2}}+\frac{1}{\sqrt{1-x^2}}\right)\right]dx,\\
f_3(\varkappa,\,\delta)&=\int\limits_0^{\min\{\frac{1+\delta}{\varkappa},\,1\}}\frac{\varkappa/\sqrt{\frac{1+\delta}{\varkappa}-x}+1/\sqrt{1-x}}
{1+\frac{x}{(1+\varkappa)^2}\left[\frac{\varkappa}{\sqrt{\frac{1+\delta}{\varkappa}-x}}+\frac{1}{\sqrt{1-x}}\right]^2}dx,
\end{split}
\end{equation*}
at $1+\delta<\varkappa$ (light doping)
\begin{equation*}
\begin{split}
f_2(\varkappa,\,\delta)&=\int\limits_{\sqrt{\frac{1+\delta}{\varkappa}}}^1\arctan\left[\frac{x}{\sqrt{1-x^2}\left(1+\varkappa-\varkappa x/\sqrt{x^2-\frac{1+\delta}{\varkappa}}\right)}\right]dx,\\
f_4(\varkappa,\,\delta)&=\int\limits_{\sqrt{\frac{1+\delta}{\varkappa}}}^1\frac{x\sqrt{1-x^2}}{(1-x^2)\left(1+\varkappa-\varkappa x/\sqrt{x^2-\frac{1+\delta}{\varkappa}}\right)^2+x^2}dx,
\end{split}
\end{equation*}
\begin{equation*}
\Theta(\varkappa,\,\delta)=\begin{cases}
\sqrt{\frac{1+\delta}{\varkappa}}+\left(1-(1+\varkappa)\sqrt{\frac{1+\delta}{\varkappa(1+2\varkappa)}}\right)
\theta\left(\varkappa-\frac{2\delta+1+\sqrt{5+4\delta}}{2(1-\delta)}\right), & \delta<1,\\
\sqrt{\frac{1+\delta}{\varkappa}}, & \delta\geq1;
\end{cases}
\end{equation*}
at $1+\delta>\varkappa$ (high doping)
\begin{equation*}
\begin{split}
f_2(\varkappa,\,\delta)&=\int\limits_1^{\sqrt{\frac{1+\delta}{\varkappa}}}\arctan\left[\frac{\varkappa x}{\sqrt{\frac{1+\delta}{\varkappa}-x^2}\left(1+\varkappa- x/\sqrt{x^2-1}\right)}\right]dx,\\
f_4(\varkappa,\,\delta)&=\int\limits_1^{\sqrt{\frac{1+\delta}{\varkappa}}}\frac{\varkappa x\sqrt{\frac{1+\delta}{\varkappa}-x^2}}{(\frac{1+\delta}{\varkappa}-x^2)\left(1+\varkappa-x/\sqrt{x^2-1}\right)^2+\varkappa^2x^2}dx,\\
\Theta(\varkappa,\,\delta)&=1+\left(\frac{1+\delta}{\varkappa}-\frac{1+\varkappa}{\sqrt{\varkappa(2+\varkappa)}}\right)
\theta\left(\frac{1}{2}\left(\delta-1+\sqrt{5+6\delta+\delta^2}\right)-\varkappa\right).
\end{split}
\end{equation*}
\end{widetext}
Here, $\theta(x)$ is the Heaviside function,
\begin{equation*}
\theta(x)=\begin{cases}
0, & x<0,\\
1, & x>0.
\end{cases}
\end{equation*}

We approximated the function $I(q)$ in the intermediate region $1\lesssim q\lesssim3$ by the line segment \cite{Andryushin1976b, Andryushin1976c}. The correlation energy is
\begin{equation}\label{Ecorr}
\begin{split}
&E_\text{corr}=\frac{1}{2}\left(-aq_1+bq^{3/2}_1-cq^2_1\right)q_2+\frac{1}{2}\left(dq_2-\frac{b}{5}\right)q^{5/2}_1\\
&+\frac{1}{2}\left(fq_2+\frac{c}{3}\right)q^3_1
-\frac{3d}{14}q^{7/2}_1-\frac{f}{4}q^4_1-\frac{g}{q^2_2}\left(1-\frac{q_1}{2q_2}\right),
\end{split}
\end{equation}
where $q_1$ and $q_2$ are the matching points of the asymptotics \eqref{Iq} with the line segment.

The point $q_1$ lies near the minimum point $q_0$ of the asymptotics $I(q)$ for $q\ll1$. For not too large $\nu_{e,h}$ ($\nu_{e, h}\leq4$) and $1\lesssim r_s\lesssim2$, the point $q_0$ is near 1
\begin{equation*}
q_1\approx q_0\approx\frac{a-\frac{3}{4}b+\frac{5}{4}d+3f}{\frac{3}{4}b-2c+\frac{15}{4}d+6f}.
\end{equation*}
We find for $\nu_{e,h}\gg1$
\begin{equation*}
q_1\approx q_0\approx2\sqrt{2}(1+\delta)^{1/4}\sqrt{\frac{1+\sqrt{\varkappa(1+\delta)}}{\varkappa^{3/2}+\sqrt{1+\delta}}}.
\end{equation*}

With a good accuracy (within 10\%) the second matching point is
\begin{equation*}
q_2\approx\left(\frac{4g}{|I(q_0)|}\right)^{1/3}.
\end{equation*}

We obtain in the multivalley case ($\nu_{e,h}\gg1$) the lower bound for the correlation energy $E_\text{corr}\gtrsim-A(\varkappa,\,\delta)n^{1/3}$, where
\begin{equation}\label{EstimEcorr}
A(\varkappa,\,\delta)=2\left(\frac{12}{\pi}\right)^{1/3}\frac{(1+\delta)^{1/6}(1+\sqrt{ \varkappa(1+\delta)})}{\left(1+\frac{\delta}{2}\right)^{1/3}(\varkappa^{3/2}+\sqrt{1+\delta})^{1/3}}.
\end{equation}

The coefficient \eqref{EstimEcorr} as a function of $\varkappa$ has a maximum at $\varkappa_\text{max}=1+\delta$ (for a given $\varkappa$, this happens when $n_e=\varkappa n$). Then we have
\begin{equation}\label{EstimEcorrmax}
E_\text{corr}\gtrsim-4\left(\frac{6}{\pi}\right)^{1/3}\left(1+\frac{\delta}{2}\right)^{1/3}n^{1/3}.
\end{equation}
Therefore, for $\delta\neq0$ ($n_e>n_h$), it is favorable to have $\varkappa\approx\varkappa_\text{max}>1$ ($\nu_e>\nu_h$). This means that in order to increase the correlation energy (in modulus), one should add those charge carriers whose number of valleys is greater.

\begin{figure*}[t!]
\begin{center}
\includegraphics[width=0.45\textwidth]{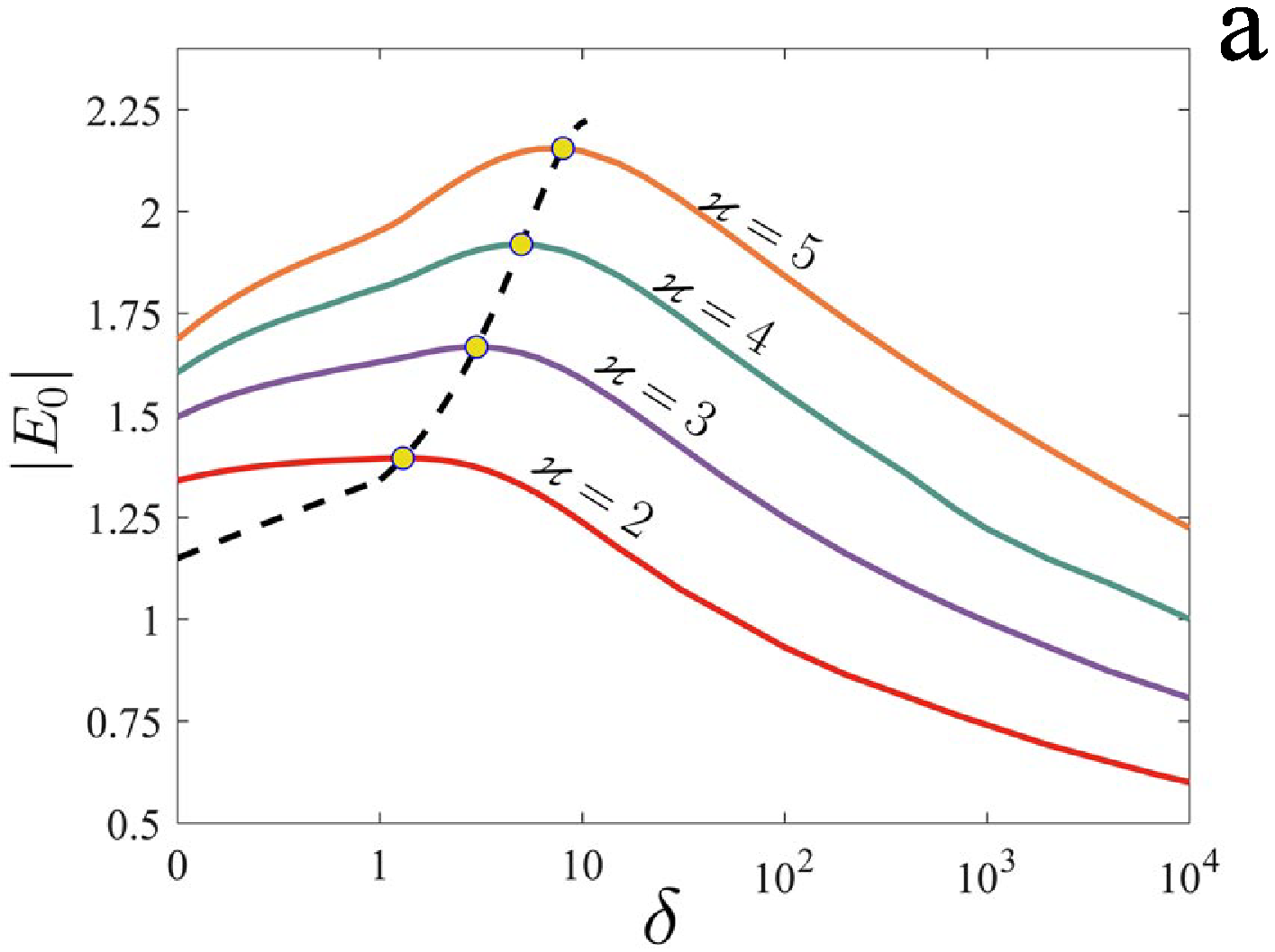}\hspace{0.05\textwidth}\includegraphics[width=0.45\textwidth]{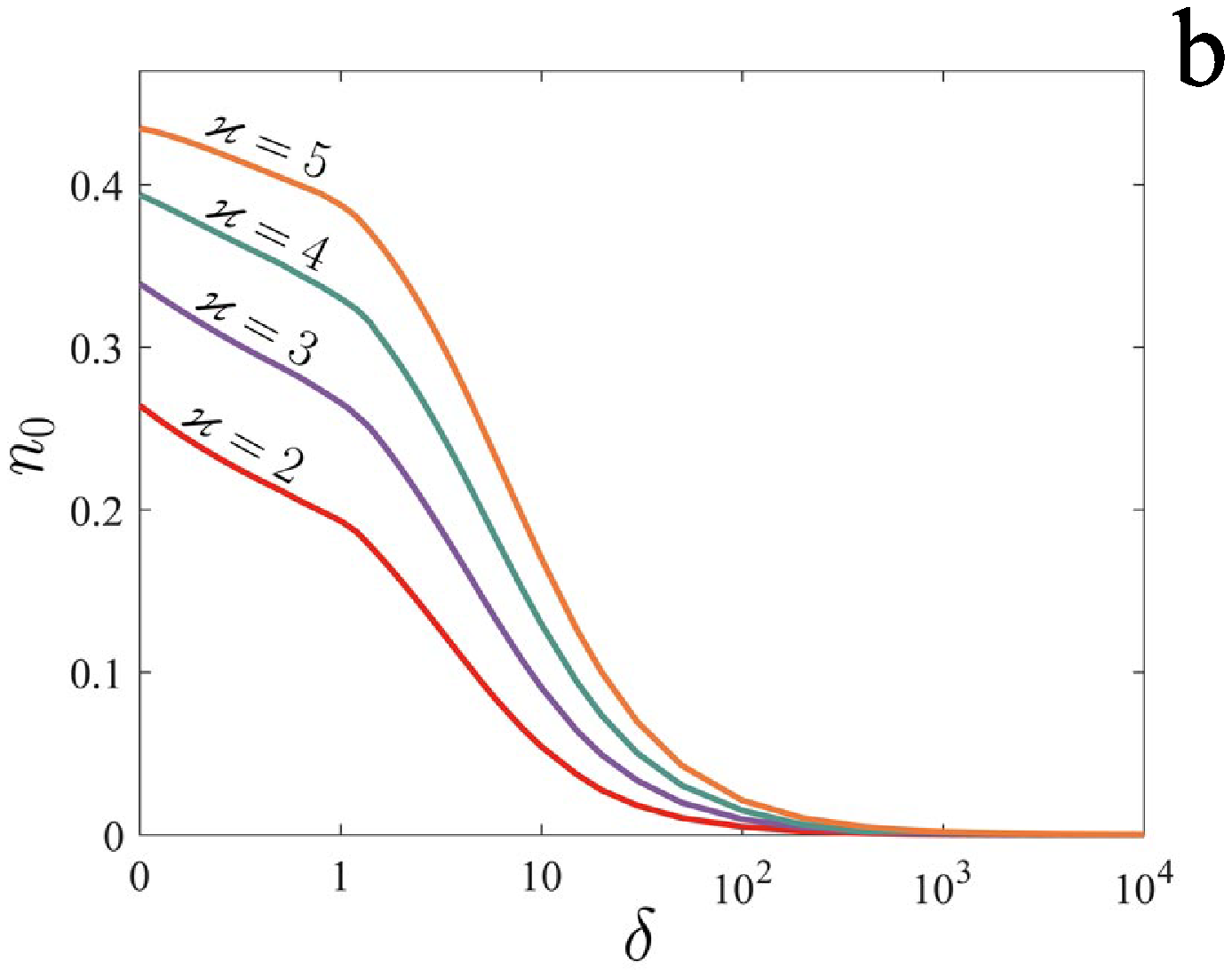}
\caption{\label{f1} Numerical calculation of the dependence of the binding energy (a) and the equilibrium density (b) of the EHL on the electron doping level $\delta$ in the case of equal masses of an electron and a hole for a different number of electron valleys $\nu_e$ and one hole valley (more precisely, two hole valleys with the removal of spin degeneracy in them). Yellow dots mark the maxima of $|E_0|$, which lie at $\delta_\text{max}=1.3,~3,~5,~8$ and agree well with the estimate \eqref{Estim_delta_max}. Black dotted line is square polynomial approximation for the position of these maxima. For $\varkappa\rightarrow1$ $\delta_\text{max}\rightarrow0$ this approximation gives $\max|E_0|\rightarrow1.15$, which is in good agreement with the EHL binding energy in an undoped single-valley semiconductor $|E_0|=1.09$ \cite{Pekh2021}.}
\end{center}
\end{figure*}

The equilibrium EHL density and energy for $\nu_{e,h}\gg1$ and $\sigma=1$ are
\begin{eqnarray*}
n_0&=&\frac{2^4}{3\pi^2}\frac{(1+\delta)^{1/4}(1+\sqrt{\varkappa(1+\delta)})^{3/2}\nu^{3/2}_h}{\left(1+\frac{\delta}{2}\right)^{1/2}(\varkappa^{3/2}+\sqrt{1+\delta})^{1/2}
\left(1+\frac{1+\delta}{\varkappa}\right)^{3/2}},\\
E_0&=&-\frac{2^4}{3\pi}\frac{(1+\delta)^{1/4}(1+\sqrt{\varkappa(1+\delta)})^{3/2}\nu^{1/2}_h}{\left(1+\frac{\delta}{2}\right)^{1/2}(\varkappa^{3/2}+\sqrt{1+\delta})^{1/2}
\left(1+\frac{1+\delta}{\varkappa}\right)^{1/2}}.
\end{eqnarray*}
We neglect the exchange energy, and for the correlation energy we use the estimate \eqref{EstimEcorr}. In the cases of $\varkappa\ll1$ and $\varkappa\gg1$ for a given $\delta$, $n_0$ and $|E_0|$ increase, i.e. EHL becomes more compressed and more tightly bound when charge carriers are added in band with larger number of valleys (this corresponds for $n_e>n_h$ to $\varkappa\gg1$)
\begin{eqnarray*}
\left.n_0\right|_{\varkappa\ll1}&=&\frac{2^4}{3\pi^2}\frac{\nu^{3/2}_e}{\left(1+\frac{\delta}{2}\right)^{1/2}(1+\delta)^{3/2}},\\
\left.E_0\right|_{\varkappa\ll1}&=&-\frac{2^4}{3\pi}\frac{\nu^{1/2}_e}{\left(1+\frac{\delta}{2}\right)^{1/2}\sqrt{1+\delta}},\\
\left.n_0\right|_{\varkappa\gg1}&=&\frac{2^4}{3\pi^2}\frac{1+\delta}{\left(1+\frac{\delta}{2}\right)^{1/2}}\nu^{3/2}_h,\\
\left.E_0\right|_{\varkappa\gg1}&=&-\frac{2^4}{3\pi}\frac{1+\delta}{\left(1+\frac{\delta}{2}\right)^{1/2}}\nu^{1/2}_h.
\end{eqnarray*}
We took into account in the first two lines that $\varkappa\nu_h=\nu_e$. Thus, for very different numbers of valleys, $n_0$ and $E_0$ are determined by a smaller number of $\nu_e$ or $\nu_h$.

Numerical calculations confirm that for intermediate values of $\varkappa$ ($0.5\lesssim\varkappa\lesssim5$), the EHL binding energy becomes larger if we add carriers with larger number of valleys. As the doping increases, $|E_0|$ increases and then begins to decrease (see Fig.~\ref{f1}a). There is an optimal value of $\delta_\text{max}\gtrsim\varkappa-1$ for a given $\varkappa>1$, when the maximum $|E_0|$ is reached. Based on the results of the performed numerical calculations, it is possible to make the estimate
\begin{equation}\label{Estim_delta_max}
\delta_\text{max}\approx\varkappa^{4/3}-1.
\end{equation}
Moreover, the equilibrium EHL density $n_0$ decreases monotonically (see Fig.~\ref{f1}b).

In the limit $\nu_{e,h}\rightarrow\infty$, the dependence of the correlation energy on $\nu_{e,h}$ and $\varkappa$ disappears
\begin{equation}\label{Ecorr_big_nu}
E_\text{corr}=-A(\sigma,\,\delta)n^{1/3},
\end{equation}
where
\begin{equation*}
\begin{split}
&A(\sigma,\,\delta)=\frac{3}{(4\pi)^{2/3}\left(1+\frac{\delta}{2}\right)}\\
&\times\int\limits_0^\infty d\xi\int\limits_0^\infty d\zeta\frac{\xi^3\left((1+\delta)\frac{\eta_e}{\xi^4+\eta^2_e\zeta^2}+\frac{\eta_h}{\xi^4+\eta^2_h\zeta^2}\right)^2}
{1+\xi\left[(1+\delta)\frac{\eta_e}{\xi^4+\eta^2_e\zeta^2}+\frac{\eta_h}{\xi^4+\eta^2_h\zeta^2}\right]}.
\end{split}
\end{equation*}

We have in a particular case $\sigma=1$
\begin{equation*}
A(1,\,\delta)=A(1)\left(1+\frac{\delta}{2}\right)^{1/3},
\end{equation*}
where $A(1)$ is the previously found (now corrected) value of the constant in the absence of doping \cite{Pekh2020}
\begin{equation*}
A(1)=\frac{3}{2^{2/3}\pi^{1/6}}\Gamma\left(\frac{2}{3}\right)\Gamma\left(\frac{5}{6}\right)\approx2.387.
\end{equation*}

The point $\sigma=1$ is the minimum point of the function $A(\sigma,\,\delta)$ at $\delta=0$ with a minimum equal to $A(1)$. When $\delta\neq0$, the function $A(\sigma,\,\delta)$ also has a minimum, but it is at $\sigma<1$, and its value is less than $A(1,\,\delta)$. The point $\sigma=1$ is located on the growing section of $A(\sigma,\,\delta)$. This means that in order to increase (in modulus) the correlation energy in the case $n_e>n_h$, it is desirable to have $\sigma>1$ ($m_e>m_h$) and vice versa for $n_e<n_h$. Therefore, when $\nu_e=\nu_h$, heavier carriers should be added for increasing the binding energy of EHL and its equilibrium density.

In the case of $\delta\gg1$, the number of holes is negligible and the correlation energy is a function of the electron density $n_e$. We normalize it to one electron
\begin{equation}\label{Ecorr_big_nu_big_delta}
E_\text{corr}=-\widetilde{A}(\sigma)n^{1/3}_e,
\end{equation}
\begin{equation*}
\widetilde{A}(\sigma)=\frac{3(1+\sigma)^{1/3}}{2^{7/3}\pi^{1/6}}\Gamma\left(\frac{2}{3}\right)\Gamma\left(\frac{5}{6}\right)\approx0.752(1+\sigma)^{1/3}.
\end{equation*}

In the region occupied by EHL, the renormalized band gap is equal to $E_g=E^{(0)}_g+E_0-E_F$, where $E^{(0)}_g$ is the initial band gap and $E_F$ is the Fermi energy of $e$-$h$ pairs \cite{Tikhodeev1985}. We have $E_0=-E_F$ for multivalley systems when the exchange energy is neglected. Therefore, the renormalization of the band gap is $\Delta E_g=2E_0$. We obtain $\Delta E_g=-\frac{4}{3}A(\sigma,\,\delta)n^{1/3}_0$ in the region of EHL in 2D multivalley systems.

In the region of $e$-$h$ plasma, the renormalized band gap is equal to $E_g=E^{(0)}_g-E_F$. If the minimum of the ground state energy is reached, we find $\Delta E_g=-\frac{2}{3}A(\sigma,\,\delta)n^{1/3}$. This corresponds to the well-established law $\Delta E_g\propto n^{1/3}$ for 2D $e$-$h$ plasmas \cite{Yu2019, Trankle1987}.

We consider the question about the band gap renormalization due to the electron (hole) gas in Sec.~\ref{s5}.

\section{\label{s3}Gas--liquid transition}

In the absence of doping ($\delta=0$), the thermodynamics of $e$-$h$ pairs in TMD monolayers was described in our previous work \cite{Pekh2021}. At low doping levels ($\delta\ll1$, $n_e\approx n_h$), the picture changes slightly. Carriers added with the help of doping participate in the formation of trions, which is energetically favorable (the binding energy of a trion in TMDs is tens of meV). In this case, the fraction of trions in comparison with excitons in the gas phase is small. In contrast to the insulating gas of excitons, the gas of trions is conductive. Qualitative changes in the kinetic (transport) properties of the system begin to manifest themselves when the number of trions becomes comparable to the number of excitons. The thermodynamics of the gas--liquid transition does not undergo significant changes.

A specified case is the equality of the number of excess charge carriers and the number of $e$-$h$ pairs ($\delta=1$). The exciton gas is completely converted into the trion gas. Quenching of exciton lines in the photoluminescence spectrum of TMD monolayers is observed at high doping levels \cite{Mak2013, Ross2013, Yang2015}. Instead of the reaction of formation (decay) of the exciton $e+h\rightleftarrows X^0$, there is the reaction of formation (decay) of the electron trion $e+X^0\rightleftarrows X^-$ or the hole trion $h+X^0\rightleftarrows X^+$ (reactions $e+e+h\rightleftarrows X^-$ and $e+h+h\rightleftarrows X^+$ are significantly more rare ones). The properties of trions were discussed in detail in the review \cite{Durnev2018}.

The chemical potential of charge carriers in doped heterostructures is represented in the form
\begin{widetext}
\begin{equation}\label{mu_general}
\begin{split}
\mu(n,\,T)=T\ln\left\{\left[\exp\left(\frac{2\pi(1+\delta)n}{\varkappa(1+\sigma)\nu_hT}\right)-1\right]\left[\exp\left(\frac{2\pi\sigma n}{(1+\sigma)\nu_hT}\right)-1\right]\right\}-\frac{2\sqrt{2}}{\sqrt{\pi\nu_h}}\left(1+\sqrt{\frac{1+\delta}{\varkappa}}\right)\sqrt{n}+\frac{\partial}{\partial n}\left(nE_\text{corr}\right).
\end{split}
\end{equation}
The first term is the chemical potential of non-interacting electron and hole. The second and third terms are, respectively, the exchange and correlation contributions, which are assumed to be independent of the temperature $T$, since their temperature corrections cancel at $T\ll E_F$~\cite{Andryushin1979}.

The critical density $n_c$ and the critical temperature $T_c$ of the gas--liquid transition are determined by the equations
\begin{equation}\label{crit_point}
\left.\frac{\partial\mu}{\partial n}\right|_{\substack{n=n_c\\ T=T_c}}=\left.\frac{\partial^2\mu}{\partial n^2}\right|_{\substack{n=n_c\\ T=T_c}}=0.
\end{equation}

To obtain the temperature dependence of the $e$-$h$ pair density in the gas and liquid phases $n_\text{G}(T)$ and $n_\text{L}(T)$, we use the Maxwell construction
\begin{equation}\label{Maxwell_rule}
\int\limits_{n_\text{G}}^{n_\text{L}}\mu(n,\,T)dn=\mu(T)(n_\text{L}-n_\text{G}),
\end{equation}
where $\mu(T)=\mu(n_\text{G},\,T)=\mu(n_\text{L},\,T)$.

Each value of $\delta$ has its bell-shaped curve formed by a set of pairs of points $n_\text{G}$ and $n_\text{L}$ on the plane $(n,\,T)$. Thus, the gas--liquid transition at doping corresponds to a surface in the space $(\delta,\,n,\,T)$. Such a surface for a system with equal masses of charge carriers, $\nu_e=2$, and $\nu_h=1$ is shown in Fig.~\ref{f2}. The ratios of equilibrium to critical parameters are weakly dependent on $\delta$ (see the inset in Fig.~\ref{f2}), therefore the dependencies of $T_c$ and $n_c$ on $\delta$ are similar to the dependencies of $|E_0|$ and $n_0$ on $\delta$ for various $\varkappa$, which are shown in Fig.~\ref{f1}. The $n_c(\delta)$ dependence is monotonically decreasing, and the $T_c(\delta)$ dependence for $\varkappa>1$ has a maximum. To estimate the position of the latter, one can be also use the formula \eqref{Estim_delta_max}.

\begin{figure*}[t!]
\begin{center}
\includegraphics[width=0.7\textwidth]{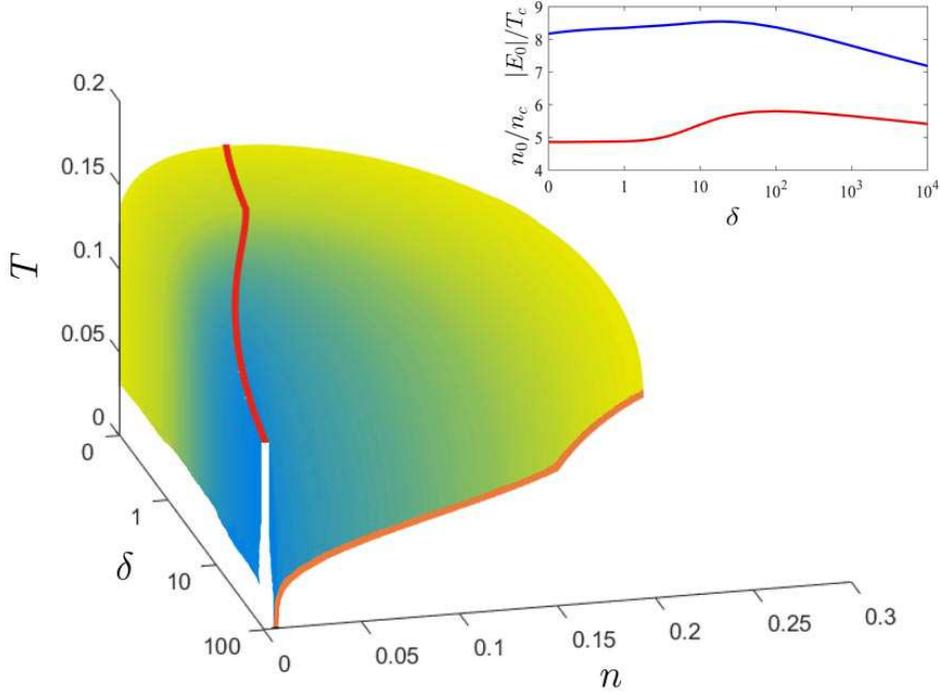}
\caption{\label{f2} 3D phase diagram $(\delta,\,n,\,T)$ of a gas--liquid transition for equal masses of charge carriers and two electron and one hole valleys (or two hole valleys, but with the removal of the spin degeneracy). The red curve on the surface is the critical points $(n_c,\,T_c)$ for the corresponding values of $\delta$. The orange curve in the $T=0$ plane is the equilibrium density $n_0$ as a function of $\delta$. The inset shows the $\delta$ dependencies of the relations $n_0/n_c$ (red curve) and $|E_0|/T_c$ (blue curve).}
\end{center}
\end{figure*}

In the multivalley case ($\nu_{e,h}\rightarrow\infty$), the exchange contribution to both the energy $E_\text{gs}$ and the chemical potential can be neglected, and we can set $\varkappa=1$ and $\nu_e=\nu_h=\nu$. Then the pair of equations \eqref{crit_point} is reduced \cite{Andryushin1977} to one equation
\begin{equation}\label{eq_zc}
z_c=(1-\gamma)\frac{\left(\cosh z_c-\cosh(s_\delta z_c)\right)\left(e^{z_c}-\cosh(s_\delta z_c)-s_\delta\sinh(s_\delta z_c)\right)}{(1+s_\delta^2)\left(\cosh z_c\cosh(s_\delta z_c)-1\right)-2s_\delta\sinh z_c\sinh(s_\delta z_c)},
\end{equation}
where the unknown
\begin{equation*}
z_c=\frac{\pi n_c}{\nu T_c}\left(1+\frac{\delta}{2}(1+s)\right)
\end{equation*}
and
\begin{equation*}
s_\delta=\frac{s+\frac{\delta}{2}(1+s)}{1+\frac{\delta}{2}(1+s)},~~~s=\frac{1-\sigma}{1+\sigma}.
\end{equation*}
Here and below, $\gamma=\hspace{0.05cm}^1\hspace{-0.08cm}/_3$ is the exponent of the density and $A=A(\sigma,\,\delta)$ is the coefficient in the correlation energy \eqref{Ecorr_big_nu}. These equations and the formulas below contain an arbitrary $\gamma$, since they also apply to layered systems with $\gamma=\hspace{0.05cm}^1\hspace{-0.08cm}/_4$. The corresponding expressions for them in the absence of doping were written in Ref. \cite{Andryushin1977}. The TMD monolayer, generally speaking, is a three-layer system, in which a layer of transition atoms is encapsulated between two layers of chalcogen atoms. For this reason, the relevant value of $\gamma$ can be intermediate between $\hspace{0.05cm}^1\hspace{-0.08cm}/_4$ and $\hspace{0.05cm}^1\hspace{-0.08cm}/_3$. The calculation of the effective $\gamma$ for the TMD monolayer with a small number of valleys can also be carried out at electron (hole) doping, following our previous work \cite{Pekh2021}. In particular, the effective $\gamma$ at $\nu_e=\nu_h=4$ turns out to be close to $\hspace{0.05cm}^1\hspace{-0.08cm}/_4$.

The critical parameters of EHL are expressed in terms of $z_c$ by the equations
\begin{eqnarray}
n_c&=&\left[\frac{\gamma(1+\gamma)\nu A}{\pi\left(1+\frac{\delta}{2}(1+s)\right)}\frac{\cosh z_c-\cosh(s_\delta z_c)}{e^{z_c}-\cosh(s_\delta z_c)-s_\delta\sinh(s_\delta z_c)}\right]^\frac{1}{1-\gamma},\\
T_c&=&\frac{\pi}{z_c}\left(\frac{\nu}{1+\frac{\delta}{2}(1+s)}\right)^{\frac{\gamma}{1-\gamma}}\left[\frac{\gamma(1+\gamma)A}{\pi}\frac{\cosh z_c-\cosh(s_\delta z_c)}{e^{z_c}-\cosh(s_\delta z_c)-s_\delta\sinh(s_\delta z_c)}\right]^\frac{1}{1-\gamma},\\
\mu_c&=&-T_c\left\{\left(\frac{1}{\gamma}-1+\frac{1}{\gamma}\frac{\sinh z_c-s_\delta\sinh(s_\delta z_c)}{\cosh z_c-\cosh(s_\delta z_c)}\right)z_c-\ln2-\ln\left(\cosh z_c-\cosh(s_\delta z_c)\right)\right\}.
\end{eqnarray}
\end{widetext}

The ratios $n_0/n_c$ and $|E_0|/T_c$ are independent of $A$ and weakly depend on $\sigma$ only through $s_\delta$
\begin{eqnarray*}
\frac{n_0}{n_c}&=&\left[\frac{e^{z_c}-\cosh(s_\delta z_c)-s_\delta\sinh(s_\delta z_c)}{(1+\gamma)\left(\cosh z_c-\cosh(s_\delta z_c)\right)}\right]^\frac{1}{1-\gamma},\\
\frac{|E_0|}{T_c}&=&\frac{1-\gamma}{\gamma}z_c\left[\frac{e^{z_c}-\cosh(s_\delta z_c)-s_\delta\sinh(s_\delta z_c)}{(1+\gamma)\left(\cosh z_c-\cosh(s_\delta z_c)\right)}\right]^\frac{1}{1-\gamma}.
\end{eqnarray*}
It follows that the ratios of the equilibrium parameters to the critical ones are similar for different systems with different $\sigma$ and $\delta$, but with the same $s_\delta$.

Expanding the equation \eqref{eq_zc} in powers of $z_c$ up to terms of the third order, we find
\begin{equation}\label{approx_zc}
z_c(s_\delta)\approx\frac{3}{2(1+s_\delta^2)}\left[\sqrt{1+\frac{8\gamma}{3(1-\gamma)}(1+s_\delta^2)}-1\right],
\end{equation}
from which we obtain the critical parameters
\begin{eqnarray}
n_c&\approx&\left[\frac{\gamma(1-\gamma^2)A\nu}{2\pi}\left(1+\frac{1+s_\delta^2}{12}z^2_c\right)\right]^\frac{1}{1-\gamma},\\
T_c&\approx&\frac{\pi\nu^{\frac{\gamma}{1-\gamma}}}{z_c}\left[\frac{\gamma(1-\gamma^2)A}{2\pi}\left(1+\frac{1+s_\delta^2}{12}z^2_c\right)\right]^\frac{1}{1-\gamma},
\end{eqnarray}
\begin{equation}
\begin{split}
\mu_c\approx-T_c&\left\{\frac{2}{\gamma(1-\gamma)}\left(1-\frac{1+s_\delta^2}{12}z^2_c\right)-z_c\right.\\
&\left.-\ln\left[(1-s_\delta^2)z^2_c\left(1+\frac{1+s_\delta^2}{12}z^2_c\right)\right]\right\}.
\end{split}
\end{equation}

Using the Maxwell construction \eqref{Maxwell_rule}, we find the temperature dependencies of the $e$-$h$ pair density in the gas and liquid phases and the chemical potential near the critical point $\frac{T_c-T}{T_c}\ll1$
\begin{equation}
\begin{split}
&n_c-n_\text{G}(T)=n_\text{L}(T)-n_c\\
&\approx\sqrt{\frac{6}{\gamma}}\left(1-\frac{1+s_\delta^2}{12\gamma}z^2_c\right)n_c\left(\frac{T_c-T}{T_c}\right)^{1/2},
\end{split}
\end{equation}
\begin{equation}
\begin{split}
&\mu(T)-\mu_c\\
=&\left\{2-\ln\left[(1-s_\delta^2)z^2_c\right]+\frac{1+s_\delta^2}{12}z^2_c\right\}\left(T_c-T\right).
\end{split}
\end{equation}

At low temperatures $T\ll T_c$, the density in the gas phase is exponentially small, while in the liquid phase it deviates from the equilibrium EHL density $n_0$ quadratically in temperature
\begin{equation}
\begin{split}
n_L(T)=&n_0\left\{1-\frac{\pi^2\left(1+\gamma\right)^{\frac{2}{1-\gamma}}}{6(1-s_\delta^2)}\left(\frac{1-\gamma}{2}\right)^{\frac{1+\gamma}{1-\gamma}}\right.\\
&\left.\times z^{\frac{2\gamma}{1-\gamma}}_c\left(1+\frac{1+s_\delta^2}{6(1-\gamma)}z^2_c\right)\left(\frac{T}{T_c}\right)^2\right\}.
\end{split}
\end{equation}
The correction to the chemical potential is also quadratic in temperature
\begin{equation}\label{mu_lowT}
\begin{split}
\mu(T)&=\mu_0\left\{1+\frac{\pi^2\gamma\left(1+\gamma\right)^{\frac{2}{1-\gamma}}}{6(1-s_\delta^2)}\left(\frac{1-\gamma}{2}\right)^{\frac{1+\gamma}{1-\gamma}}\right.\\
&\left.\times z^{\frac{2\gamma}{1-\gamma}}_c\left(1+\frac{1+s_\delta^2}{6(1-\gamma)}z^2_c\right)\left(\frac{T}{T_c}\right)^2\right\}.
\end{split}
\end{equation}

\section{\label{s4}Trion gas--electron-hole plasma transition}

As the number of $e$-$h$ pairs increases, a transition from the insulating state of the system (exciton gas) to the metallic one (electron-hole plasma) was observed without doping. This is a metal--insulator transition (Mott transition). The physical reason for this phenomenon is in the screening of the Coulomb interaction between charge carriers with increasing the density. At the point of the transition, the exciton binding energy vanishes and the exciton decays into free electron and hole.

If the temperature is significantly lower than the binding energy of a trion, 20-30 meV \cite{Mak2013, Ross2013, Yang2015}, there is a metallic phase in which trions are free charge carriers. Trion conduction has not been studied and may have unusual properties.

Taking into account the possibility of achieving high densities of excess charge carriers in 2D materials, in order to avoid confusion, it would be appropriate to speak about the transition from a trion gas to an electron-hole plasma rather than the metal--insulator transition. It occurs for the same physical reason as the Mott transition. As the density increases, the edge of the continuous spectrum shifts faster than the energy of the bound state. At the density of the transition, the bound state goes over into a continuous spectrum and it ceases to exist \cite{Kraeft1986}.

There are three possible experimental realizations of the trion gas--electron-hole plasma transition.

Firstly, one can fix the voltage at the gate and the density of excess electrons (holes). At the same time, we increase the intensity of the exciting laser radiation, i.e. increase the density of generated $e$-$h$ pairs $n$. Then the density of the transition will depend on the initial doping level $\delta_0$. We expect it to fall off as $\delta_0$ increases.

Secondly, one can fix the intensity of laser radiation $I_0$ and increase the density of excess electrons or holes. In this case, the excess charge carriers screen the trions. The doping level increases, and the required value $\delta_c$ for the transition decreases with increasing $I_0$.

Thirdly, it is possible to maintain a certain ratio between the $e$-$h$ pair density $n$ and the excess carrier density $\Delta n$, i.e. one can fix the doping level $\delta$, increasing parallel both the radiation intensity and the gate voltage. As noted above, an interesting case is a pure trion gas, when $\delta=1$. Then the evolution of the trion gas can be investigated, although this is experimentally difficult to implement. At the density of the transition $n_\text{tr}$, trions undergo decay $X^-\rightarrow X^0+e\rightarrow e+h+e$ ($X^+\rightarrow X^0+h\rightarrow e+h+h$). Due to the stronger screening of the Coulomb interaction in the presence of excess charge carriers, it turns out to be less than the density of the Mott transition in the absence of doping, $n_\text{tr}<n_\text{dm}$.

Below we adhere to the third formulation of the problem as the most interesting from a theoretical point of view. Although its experimental implementation is extremely difficult.

In the range of densities above $n_\text{tr}$, the system behaves like 3D doped semiconductors and free electrons and holes are charge carriers. In our case, the system is a semi-metallic, since it contains both electrons and holes.

\subsection{\label{s4a}Modified Mott criterion}

The density of the metal--insulator transition in 3D semiconductors can be obtained based on the Mott criterion \cite{Mott1968, Mott1972, Mott1973}
\begin{equation}\label{Mott_criterion_3D}
n^{1/3}_\text{3D}a^\text{(3D)}_B=C.
\end{equation}
$a^\text{(3D)}_B$ is the Bohr radius of 3D exciton. For a number of traditional semiconductors, the constant $C=0.26\pm0.05$ \cite{Edwards1978}.

This criterion is justified only for 3D materials, when the Debye screening of the Coulomb interaction between charge carriers is exponential. For 2D materials, screening is not exponential one. Charge carriers are in the film and interact mainly through its environment. However, for simple quantitative estimates of the density of the metal--insulator transition in 2D materials, one can use the Mott criterion \cite{Courtade2017, Glazov2018}, written in the form
\begin{equation}\label{Mott_criterion_2D}
na^2_B=\widetilde{C}.
\end{equation}
$a_B$ is the Bohr radius of 2D exciton.

By squaring both sides of \eqref{Mott_criterion_3D} and expressing $a^\text{(3D)}_B$ in terms of $a_B$ as $a_B=\xi a^\text{(3D)}_B$, one can get an estimate
\begin{equation}\label{Const_Mott_2D}
\widetilde{C}\simeq\xi^2C^2.
\end{equation}
Coefficient $\xi\geq\hspace{0.05cm}^1\hspace{-0.08cm}/_2$. It is equal to $\hspace{0.05cm}^1\hspace{-0.08cm}/_2$ for the 2D Coulomb potential
\begin{equation}\label{Coulomb_potential_2D}
V(q)=\frac{2\pi\widetilde{e}^2}{q}.
\end{equation}
Here, as above, $\widetilde{e}^2=e^2/\varepsilon_\text{eff}$ and $\varepsilon_\text{eff}=(\varepsilon_1+\varepsilon_2)/2$, $\varepsilon_{1,2}$ are permittivities of the surrounding media.

The interaction of charge carriers in thin films is determined by the Keldysh potential \cite{Rytova1967, Chaplik1971, Keldysh1979}
\begin{equation}\label{Keldysh_int}
V_K(q)=\frac{2\pi\widetilde{e}^2}{q(1+r_0q)},
\end{equation}
where $r_0$ is the screening length, $r_0=d\varepsilon/(\varepsilon_1+\varepsilon_2)$, $d$ is the film thickness, $\varepsilon$ is the permittivity of the film material, $\varepsilon\gg\varepsilon_{1,2}$. As applied to systems with TMD monolayers, the value $r_0$ is an adjustable parameter that was used to calculate excitons \cite{Durnev2018}.

The Keldysh potential is weaker than the potential \eqref{Coulomb_potential_2D}, therefore $\xi>\hspace{0.05cm}^1\hspace{-0.08cm}/_2$ for excitons in TMDs. The Bohr radius of the exciton is found by the variational solution of the Schr\"{o}dinger equation with the potential \eqref{Keldysh_int} as the variational parameter, which minimizes the energy. Usually $\xi\approx1$, but it may be more than 1, since $\varepsilon\gg\varepsilon_{1,2}$ and $r_0$ can be so large that the Keldysh potential turns out to be weaker than the 3D Coulomb potential with the effective permittivity of the medium $\varepsilon_\text{eff}$. It turns out $a_B\gtrsim a^\text{(3D)}_B$. In our opinion, this result can be a sufficient justification for the estimate \eqref{Const_Mott_2D}.

Experiments confirm that the constant $\widetilde{C}<1$ \cite{Chernikov2015b}. Theoretical estimates give $\widetilde{C}\simeq0.1$ \cite{Glazov2018}.

We propose to adapt the Mott criterion to the trion gas--electron-hole plasma transition in heterostructures based on monolayer or bilayer TMD films. For trions, one needs to calculate $n_\text{tr}a^2_\text{tr}$, where $n_\text{tr}$ is the trion density at the transition and $a_\text{tr}$ is the trion radius. If we take into account that $a_\text{tr}\gtrsim a_B$, we still get $n_\text{tr}a^2_\text{tr}<1$.

Our calculations of the metal--insulator transition \cite{Pekh2021} showed that the density of the transition $n_\text{dm}$ has a noticeable temperature dependence. $n_\text{dm}$ monotonically increases with $T$. Since the physical mechanism of screening of the Coulomb interaction is the same for excitons and trions, using our previous calculations $n_\text{dm}$, we can write the Mott criterion for the trion gas--electron-hole plasma transition in the form
\begin{equation}\label{Mott_criterion_2D_trion}
n_\text{tr}a^2_\text{tr}=\zeta(T).
\end{equation}
The function $\zeta(T)$ is approximated with a good accuracy on the temperature range $\frac{1}{4}T_c\lesssim T\lesssim T_c$ by the quadratic function
\begin{equation}\label{zeta_T}
\zeta(T)\approx\alpha T^2+\beta T+\zeta_0.
\end{equation}
This approximation is not valid at low temperatures ($T<\frac{1}{4}T_c$). We are interested in the pointed out above temperature range, since the temperature dependence of the density of the transition under consideration in this range allows us to determine in which phase (gas or liquid) it occurs. If $n_\text{tr}(T\simeq T_c)<n_c$, the transition occurs in the gas phase. If $n_\text{tr}(T\simeq T_c)>n_c$, the transition occurs in the liquid phase.

The functions $\zeta(T)$ for excitons and trions can be considered rather close at low temperatures, therefore the differences in the densities $n_\text{tr}$ and $n_\text{dm}$ are mainly due to the difference in $a_\text{tr}$ from $a_B$
\begin{equation}\label{trion_density}
n_\text{tr}\approx\left(\frac{a_B}{a_\text{tr}}\right)^2n_\text{dm}.
\end{equation}
Hence $n_\text{tr}\lesssim n_\text{dm}$. This conclusion is confirmed by the results of variational calculations provided below.

\begin{table*}
\caption{\label{t1} The reduced mass of electron and hole, the effective permittivity, the screening length, the Bohr radius of 2D exciton, the trion radius, and squared ratio of the exciton and trion radii. $m_0$ is the free electron mass.}
\begin{ruledtabular}
\begin{tabular}{cccccccc}
№&Heterostructure&$m$ ($m_0$)&$\varepsilon_\text{eff}$& $r_0$ (\AA)&$a_B$ (\AA)&$a_\text{tr}$ (\AA)&$\left(\frac{a_B}{a_\text{tr}}\right)^2$ \\
\hline
1&MoS$_2$/SiO$_2$ & 0.32$\pm$0.04 \cite{Eknapakul2014} & 2.45 & 16.926 \cite{Berkelbach2013} & 10.003 & 13.861 & 0.521\\
2&MoSe$_2$/SiO$_2$ &0.283 \cite{Rasmussen2015} & 2.45 & 21.106 \cite{Berkelbach2013} & 11.715 & 16.230 & 0.521\\
3&WS$_2$/SiO$_2$ & 0.22 \cite{Rasmussen2015} & 2.45 & 15.464 \cite{Berkelbach2013} & 11.942 & 16.517 & 0.522\\
4&WSe$_2$/SiO$_2$ & 0.23 \cite{Rasmussen2015} & 2.45 & 18.414 \cite{Berkelbach2013} & 12.494 & 17.291 & 0.522\\
5&$h$BN/MoS$_2$/$h$BN & 0.275$\pm0.015$ \cite{Goryca2019} & 4.45 & 7.640 \cite{Goryca2019} & 11.426 & 15.788 & 0.524\\
6&$h$BN/MoSe$_2$/$h$BN & 0.35$\pm0.015$ \cite{Goryca2019} & 4.4 & 8.864 \cite{Goryca2019} & 10.321 & 14.254 & 0.524\\
7&$h$BN/MoTe$_2$/$h$BN & 0.36$\pm0.04$ \cite{Goryca2019} & 4.4 & 14.546 \cite{Goryca2019} & 12.301 & 17.012 & 0.523\\
8&$h$BN/WS$_2$/$h$BN & 0.175$\pm$0.007 \cite{Goryca2019} & 4.35 & 7.816 \cite{Goryca2019} & 15.293 & 21.145 & 0.523\\
9&$h$BN/WSe$_2$/$h$BN & 0.2$\pm$0.01 \cite{Goryca2019} & 4.5 & 10 \cite{Goryca2019} & 15.598 & 21.555 & 0.524\\
\end{tabular}
\end{ruledtabular}
\end{table*}

The density of the transition $n_\text{tr}$ is obtained using the modified Mott criterion \eqref{Mott_criterion_2D_trion} with the approximation \eqref{zeta_T} if the trion radius $a_\text{tr}$ is known. The latter is found by the variational solution of the three-particle Coulomb problem with the trion Hamiltonian \cite{Courtade2017}
\begin{equation}\label{Hamiltonian_trion}
\begin{split}
\widehat{H}_\text{tr}=&-\frac{1}{2m}\left(\triangle_1+\triangle_2\right)-\frac{1}{m_{ni}}{\boldsymbol\nabla}_1{\boldsymbol\nabla}_2\\
&-V_K(\rho_1)-V_K(\rho_2)+V_K(|{\boldsymbol\rho}_1-{\boldsymbol\rho}_2|),
\end{split}
\end{equation}
where the Laplacians $\triangle_{1,2}$ and gradient operators ${\boldsymbol\nabla}_{1,2}$ act on 2D vectors of relative motion of two identical particles ${\boldsymbol\rho}_{1,2}={\boldsymbol\rho}_{i1,2}-{\boldsymbol\rho}_{ni}$ (${\boldsymbol\rho}_{i1,2}$ are radius vectors of identical particles, ${\boldsymbol\rho}_{ni}$ is radius vector of nonidentical particle; in trion $X^-$, the identical particles are electrons, the nonidentical particle is a hole and vice versa for trion $X^+$); here, as above, $m=m_em_h/(m_e+m_h)$ is the reduced mass of an electron and a hole, $m_{ni}$ is the mass of the nonidentical particle ($m_{ni}=m_h$ for trion $X^-$, $m_{ni}=m_e$ for trion $X^+$); $V_K(\rho)$ is the Keldysh potential in the coordinate space,
\begin{equation}\label{Keldysh_int_r-space}
V_K(\rho)=\frac{\pi\widetilde{e}^2}{2r_0}\left[H_0\left(\frac{\rho}{r_0}\right)-Y_0\left(\frac{\rho}{r_0}\right)\right],
\end{equation}
where $H_0$ and $Y_0$ are the Struve and Neumann functions, respectively.

At $m_e\approx m_h$, the trial wave function is simplified \cite{Berkelbach2013}
\begin{equation}\label{trial_function_r-space}
\widetilde{\psi}(\rho_1,\,\rho_2)\propto\left[\exp\left(-\frac{\rho_1}{a_1}-\frac{\rho_2}{a_2}\right)\pm\exp\left(-\frac{\rho_2}{a_1}-\frac{\rho_1}{a_2}\right)\right].
\end{equation}
Plus (minus) corresponds to a symmetric (antisymmetric) trion. The values of the variational parameters $a_{1,2}$ corresponding to the minimum energy of the trion are the effective localization radii.

As one can easily see from the view of the trial function \eqref{trial_function_r-space}, the average distance is the same for both identical particles, $\langle\rho_1\rangle=\langle\rho_2\rangle$. We define the trion radius as $a_\text{tr}=\langle\rho_{1,2}\rangle$, then
\begin{equation}\label{a_tr}
a_\text{tr}=\frac{(a_1+a_2)^6\pm64a_1^3a_2^3}{2(a_1+a_2)\left[(a_1+a_2)^4\pm16a_1^2a_2^2\right]}.
\end{equation}
If $m_e=m_h$, we obtain $a_1=a_2=a$ and $a_\text{tr}=a$ for the symmetric trion and $a_\text{tr}=1.5a$ for the antisymmetric trion. It follows that the latter has a lower binding energy than the former, which is consistent with the results of the work \cite{Courtade2017}.

We calculated the exciton and trion radii for a number of heterostructures (see Table~\ref{t1}). We given the parameters for symmetric trions, because the absolute value of the energy of antisymmetric trions is less than the exciton binding energy and they do not exist. This happens for the mass ratio $\sigma\gtrsim0.5$~\cite{Courtade2017}. Moreover, a remarkable result is that according to the estimate \eqref{trion_density} the transition density is about half the Mott transition density for all heterostructures, as $(\frac{a_B}{a_\text{tr}})^2\simeq\frac{1}{2}$.

\subsection{Variational calculation with screened potential}

As mentioned at the beginning of this section, the transition under consideration has two stages. Trions first decay into excitons and excess carriers $X^-\rightarrow X^0+e$ or $X^+\rightarrow X^0+h$. At zero temperature, this happens when the energies of the trion and exciton become equal $E_\text{tr}=E_\text{ex}$ (the trion binding energy vanishes). At a finite temperature, this occurs when the trion binding energy becomes comparable to the temperature and the trions undergo thermal dissociation
\begin{equation}\label{Criterion_transition_stage1}
E_\text{ex}-E_\text{tr}=\kappa T.
\end{equation}
By analogy with excitons due to the entropy factor \cite{Zipfel2020}, this process occurs at temperatures noticeably lower than the trion binding energy, therefore the coefficient $\kappa$ on the right side is greater than one, $1\lesssim\kappa\lesssim2$. The inaccuracy in its value has a little effect on the results of the numerical calculation and we can put $\kappa\approx\hspace{0.03cm}^3\hspace{-0.1cm}/_2$.

With a further increase in the density, excitons also undergo thermal dissociation $X^0\rightarrow e+h$ (the second stage of the transition)
\begin{equation}\label{Criterion_transition_stage2}
|E_\text{ex}|=\kappa T.
\end{equation}

The screened Coulomb potential at a high carrier density is easily found in the momentum space \cite{Pekh2021, Andryushin1979, Andryushin1981, Silin1988}
\begin{equation}
\overline{V}(\mathbf{q})=\frac{V_0(\mathbf{q})}{1+f(\mathbf{q})V_0(\mathbf{q})\Pi_0(\mathbf{q})},
\end{equation}
where the initial unscreened Coulomb interaction $V_0(\mathbf{q})$ is taken in the form of the Keldysh potential \eqref{Keldysh_int}, $\Pi_0(\mathbf{q})$ is the static 2D polarization operator of electrons and holes
\begin{equation}\label{Polarization_Operator}
\Pi_0(\mathbf{q})=\nu_e\Pi^e_0(\mathbf{q})+\nu_h\Pi^h_0(\mathbf{q}),
\end{equation}
\begin{equation*}
\Pi^{e,h}_0(\mathbf{q})=\frac{m_{e,h}}{\pi}\left\{1-\sqrt{1-\left(\frac{2q^{e,h}_F}{q}\right)^2}\theta\left(q-2q^{e,h}_F\right)\right\}.
\end{equation*}

The function $f(\mathbf{q})$ is the Hubbard correction to RPA taking into account the contribution of the exchange diagrams at high momenta \cite{Andryushin1979, Andryushin1981, Hubbard1957}. Without doping $n_e=n_h$ in the case of an equal number of valleys $\nu_e=\nu_h=\nu\geq1$, $q^e_F=q^h_F=q_F$ and
\begin{equation}\label{General_Hubb_Correct1}
f(\mathbf{q})=1-\frac{1}{4\nu}\frac{q}{q+q_F}.
\end{equation}
Here, it is taken into account that the number of the exchange diagrams increases linearly in $\nu$, while the number of the loop diagrams increases quadratically in $\nu$. Therefore, the relative contribution of the former decreases as $1/\nu$.

In the case of $\nu_e\neq\nu_h$ and/or $n_e\neq n_h$, the Fermi momenta of electrons and holes are different. It is convenient to take their geometric mean $\overline{q}_F=\sqrt{q^e_Fq^h_F}$. Then we have
\begin{equation}\label{General_Hubb_Correct2}
f(\mathbf{q})=1-\frac{1}{2(\nu_e+\nu_h)}\frac{q}{q+\overline{q}_F}.
\end{equation}

To find the dependence $E_\text{ex}(n)$, we solve by the variational method the two-particle Schr\"{o}dinger equation in the momentum space
\begin{equation}\label{Schrodinger_exciton_momentum_space}
\frac{p^2}{2m}\psi(\mathbf{p})-\int\frac{d^2q}{(2\pi)^2}\overline{V}(\mathbf{q})\psi(\mathbf{p}-\mathbf{q})=E_\text{ex}\psi(\mathbf{p}).
\end{equation}

The trial wave function for exciton is chosen in the form of the Fourier transform of the exponentially decreasing wave function
\begin{equation}\label{Psi_trial}
\widetilde{\psi}(\mathbf{p})=\frac{\sqrt{8\pi}b^2}{(b^2+p^2)^{3/2}}.
\end{equation}

To find the dependence $E_\text{tr}(n)$, we solve variationally the three-particle Schr\"{o}dinger equation
\begin{equation}\label{Schrodinger_trion_momentum_space}
\begin{split}
&\left[\frac{p_1^2+p_2^2}{2m}+\frac{1}{m_{ni}}\mathbf{p}_1\mathbf{p}_2\right]\psi(\mathbf{p}_1,\,\mathbf{p}_2)\\
&-\int\frac{d^2q}{(2\pi)^2}\overline{V}(\mathbf{q})\left[\psi(\mathbf{p}_1-\mathbf{q},\,\mathbf{p}_2)+\psi(\mathbf{p}_1,\,\mathbf{p}_2-\mathbf{q})\right]\\
&+\int\frac{d^2q}{(2\pi)^2}\overline{V}(\mathbf{q})\psi(\mathbf{p}_1-\mathbf{q},\,\mathbf{p}_2+\mathbf{q})=E_\text{tr}\psi(\mathbf{p}_1,\,\mathbf{p}_2).
\end{split}
\end{equation}

We choose the trial wave function for trion in the form of the Fourier transform of the function \eqref{trial_function_r-space}
\begin{equation}\label{trial_function_p-space}
\begin{split}
&\widetilde{\psi}(\mathbf{p}_1,\,\mathbf{p}_2)=8\pi b^2_1b^2_2\left[\left(b^2_1+p^2_1\right)^{-3/2}\left(b^2_2+p^2_2\right)^{-3/2}\right.\\
&\left.\pm\left(b^2_1+p^2_2\right)^{-3/2}\left(b^2_2+p^2_1\right)^{-3/2}\right].
\end{split}
\end{equation}
The variation parameters are related to those entered above as $b_{1,2}=a^{-1}_{1,2}$.

\begin{figure}[t!]
\begin{center}
\includegraphics[width=0.5\textwidth]{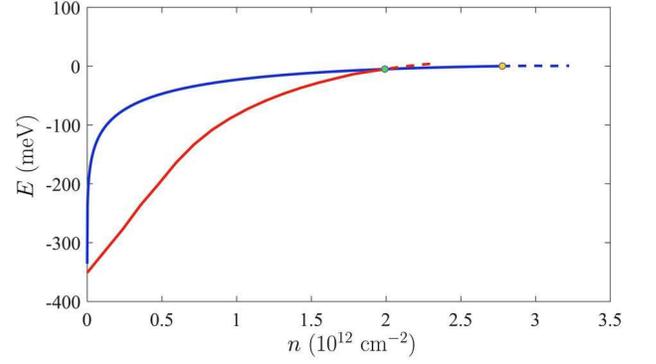}
\caption{\label{f3} Density dependence of the exciton binding energy (blue curve) and trion binding energy (red curve) in heterostructure MoS$_2$/SiO$_2$. The green point indicates the decay of trions. The yellow point indicates the decay of excitons.}
\end{center}
\end{figure}

Numerical calculation of the dependencies of the exciton and trion energies on the density confirms that the transition proceeds in two stages already at zero temperature. Fig.~\ref{f3} shows a typical behavior of $E_\text{ex}(n)$ and $E_\text{tr}(n)$ for the heterostructure MoS$_2$/SiO$_2$ as an example.

In the calculations at a finite temperature, the polarization operator was substituted in \eqref{Polarization_Operator} as the integral
\begin{equation}
\Pi^\alpha_0(\mathbf{q},\,T)=2\text{-}\text{-}\hspace{-0.357cm}\int\frac{d^2k}{(2\pi)^2}\frac{n^\alpha_F(\mathbf{k})-n^\alpha_F(\mathbf{k}+\mathbf{q})}
{\varepsilon^\alpha_{\mathbf{k}+\mathbf{q}}-\varepsilon^\alpha_\mathbf{k}},
\end{equation}
where $n^\alpha_F(\mathbf{p})=1/[\exp((\varepsilon^\alpha_\mathbf{p}-\mu^\alpha_\text{kin})/T)+1]$ is the Fermi distribution function of the $\alpha$th kind particles ($\alpha=e,\,h$) with the chemical potential $\mu^\alpha_\text{kin}$ (calculated for a gas of non-interacting particles), $\varepsilon^\alpha_\mathbf{p}=p^2/\eta_\alpha$ is the dispersion law of the $\alpha$th kind particles (plus for electrons, minus for holes). The bar near the integral means that it is taken in the sense of the principal value.

\begin{table}[b!]
\caption{\label{t2} Constants for the temperature dependency of the trion gas--electron-hole plasma transition density $n_\text{tr}(T)$ and the ratio of $n_\text{tr}=n_\text{tr}(T=0)$ and the metal--insulator transition density $n_\text{dm}$.}
\begin{ruledtabular}
\begin{tabular}{cccccc}
№&$\alpha$&$\beta$&$\zeta_0$& $\zeta(0)$ & $\frac{n_\text{tr}}{n_\text{dm}}$ \\
\hline
1 & 2.141661 & -0.616313 & 0.031726 & 0.053310 & 0.656888 \\
2 & 3.169341 & -0.701013 & 0.032581 & 0.056160 & 0.692247 \\
3 & 1.561552 & -0.492523 & 0.032815 & 0.054030 & 0.658792 \\
4 & 2.164500 & -0.569959 & 0.032549 & 0.053822 & 0.659217 \\
5 & 0.389394 & -0.252615 & 0.033263 & 0.055197 & 0.656693 \\
6 & 0.644499 & -0.329378 & 0.033792 & 0.057648 & 0.692128 \\
7 & 1.238440 & -0.440202 & 0.032775 & 0.055168 & 0.666533 \\
8 & 0.259519 & -0.210749 & 0.034000 & 0.056024 & 0.658581 \\
9 & 0.397398 & -0.256719 & 0.034074 & 0.055604 & 0.659038 \\
\end{tabular}
\end{ruledtabular}
\end{table}

The results of numerical calculations of the constants appearing in approximation \eqref{zeta_T} are presented in Table~\ref{t2} for the same heterostructures as in Table~\ref{t1} (their numbering in Table~\ref{t2} corresponds to Table~\ref{t1}). The trion radius was calculated as $a_\text{tr}=b^{-1}_0|_{n\rightarrow0}$, where $b_0=b_1=b_2$ is the value of the variational parameters at which the minimum of the trion energy is reached as a function of $b_1$ and $b_2$. For comparison, the penultimate column of Table~\ref{t2} shows the value of $\zeta$ at zero temperature, which deviates significantly from $\zeta_0$. This circumstance indicates the inapplicability of this approximation in the range of low temperatures, which was mentioned after formula \eqref{zeta_T}. The last column contains the ratio of the trion gas--electron-hole plasma transition density at zero temperature and the metal--insulator transition density. We can see that $\frac{n_\text{tr}}{n_\text{dm}}\approx\frac{2}{3}$. This result refines the result obtained in the end of Sec.~\ref{s4a}.

\section{\label{s5}Metal--insulator transition in equilibrium case}

In semiconductors and semimetals with extremely strong anisotropy of the electronic spectrum, in particular, multivalley ones, the role of correlation effects in $e$-$h$ plasma turns out to be anomalously large \cite{Andryushin1976a}. A significant decrease in energy associated with electron-electron correlations leads to self-compression of such a plasma, i.e. to the formation of EHL with a density $n_0\gg a^{-2}_x$ and a binding energy $|E_0|\gg E_x$. The appearance of EHL at sufficiently low temperatures occurs through a first-order phase transition, when the carrier density reaches a certain critical value, the saturated vapor density (see Sec.~\ref{s3}). This situation can be realized under thermodynamic equilibrium conditions for a semiconductor with a sufficiently narrow band gap $E_g$ with increasing temperature or decreasing $E_g$ \cite{Andryushin1977}.

The formation of EHL means a jump-like decrease in the band gap $E_g$ to some negative value, which corresponds to band overlap. There is a transition of the original semiconductor to a semimetal \cite{Andryushin1977}.

The metal--insulator transition was considered in detail in the work \cite{Andryushin1977} for layered semiconductors. It was also investigated in 3D TMD TiS$_2$ and TiSe$_2$ \cite{Silin1978} and in doped layered multicomponent semiconductors \cite{Andryushin1990b}.

An appreciable band gap renormalization in atomically thin WS$_2$ layers was observed in the work \cite{Chernikov2015b}.

The experimentally observed change in the temperature dependence of conductivity is considered under the metal--insulator transition in 3D traditional doped semiconductors. As the temperature rises, the conductivity of semiconductors increases, while that of metals decreases.

Doped heterostructures based on 2D materials (including monolayers or bilayers of TMDs) compare favorably with the 3D systems. The doping level in heterostructures is easily changed by the gate voltage.

The renormalization of the band gap is determined by the ground state energy of an electron gas $E_\text{gs}(n_e)$ (for definiteness, we consider electron doping)
\begin{equation}\label{renorm_band}
E_g(n_e)=E^{(0)}_g-|E^e_\text{gs}(n_e)|.
\end{equation}

The metal--insulator transition in equilibrium case occurs when the renormalized band gap becomes equal to the exciton binding energy at the electron gas density
\begin{equation}\label{Criterion_equilibrium_case}
E_g(n_e)=|E_\text{ex}(n_e)|.
\end{equation}

The ground state energy of the electron gas is determined like in the $e$-$h$ system (see Sec.~\ref{s2})
\begin{equation}\label{E_gs_electron_gas}
E^e_\text{gs}=\frac{1}{\eta_er^2_s}-\frac{4\sqrt{2}}{3\pi r_s}+\int\limits_0^\infty I(q)dq.
\end{equation}
Here, the dimensionless distance between the particles is defined as
\begin{equation*}
r_s=\sqrt{\frac{\nu_e}{\pi n_e}}.
\end{equation*}
Wave vectors are measured in Fermi wave vectors of electrons $q^e_F=\sqrt{2}/r_s$. The energy and density are measured in units of $E_x$ and $a^{-2}_x$, respectively.

The last term in \eqref{E_gs_electron_gas} is the correlation energy which is determined by the function $I(q)$
\begin{equation}\label{Iq_el_gas}
I(q)=\begin{cases}
-\frac{\sqrt{2}}{\pi r_s}q+\frac{2^{1/4}}{r^{3/2}_s\nu^{1/2}_e\eta^{1/2}_e}q^{3/2}-\frac{2(\pi-1)}{\pi r^2_s\nu_e\eta_e}q^2 & \\
+\frac{3}{2^{5/4}r^{5/2}_s\nu^{3/2}_e\eta^{3/2}_e}q^{5/2}+\frac{r^2_s\nu^2_e\eta^2_e-16}{12\sqrt{2}\pi r^3_s\nu^2_e\eta^2_e}q^3, & q\ll1,\\
-\eta_e\left(\nu_e-\frac{1}{2}\right)q^{-3},& q\gg1.
\end{cases}
\end{equation}

In the intermediate region $1\lesssim q\lesssim3$ we smoothly match the asymptotics \eqref{Iq_el_gas} by a line segment \cite{Andryushin1976b, Andryushin1976c}. The correlation energy is given by the formula \eqref{Ecorr} with the matching points $q_1$ and $q_2$ by substituting the corresponding coefficients from the asymptotics \eqref{Iq_el_gas}.

In the multivalley case ($\nu_e\gg1$), $q_1\approx2\sqrt{2}$ and $q_2\approx\left(3\pi r_s\nu_e\eta_e/2\right)^{1/3}$. Whence we get the lower bound for the correlation energy
\begin{equation}\label{EstimEcorr_el_gas}
E^e_\text{corr}\gtrsim-\left(\frac{12}{\pi}\right)^{1/3}\eta^{1/3}_en^{1/3}_e.
\end{equation}

Let be analyze the possibility of the metal--insulator transition in the equilibrium case in the model 2D multivalley semiconductor at zero temperature. We use the estimate for the correlation energy \eqref{EstimEcorr_el_gas} and neglect the contribution of the exchange energy to the ground state energy of the electron gas \eqref{E_gs_electron_gas}. Then the density corresponding to the minimum of $E_\text{gs}$ is
\begin{equation}\label{n0_el_gas}
n_{e0}=\frac{2}{3\pi^2}\eta^2_e\nu^{3/2}_e
\end{equation}
and this minimum is equal to
\begin{equation}\label{E0_el_gas}
E^e_0=-\frac{4}{3\pi}\eta_e\nu^{1/2}_e.
\end{equation}

To achieve the maximum effect of the band gap renormalization, we consider the region near the minimum of the ground state energy. For a large number of valleys, the density $n_{e0}\propto\nu^{3/2}_e$ is high and the exciton binding energy at such a density turns out to be negligible. Then the criterion for the transition \eqref{Criterion_equilibrium_case} is rewritten as
\begin{equation}\label{Criterion_estimation1}
E^{(0)}_g\approx|E^e_0|.
\end{equation}

Since the estimate of the correlation energy \eqref{EstimEcorr_el_gas} is a lower bound, and the contribution of the exchange energy is small, we can also consider \eqref{E0_el_gas} as a lower bound for the electron gas energy. This means that the criterion for the transition is represented in the form of the inequality
\begin{equation}\label{Criterion_estimation2}
E^{(0)}_g\lesssim\frac{4}{3\pi}\eta_e\nu^{1/2}_e,
\end{equation}
or the same inequality can be understood as a lower limit on the number of valleys
\begin{equation}\label{Criterion_estimation3}
\nu_e\gtrsim\left(\frac{3\pi E^{(0)}_g}{4\eta_e}\right)^2.
\end{equation}

We can estimate the characteristic value of the quantity on the right-hand side of the inequality \eqref{Criterion_estimation3} using parameters of the heterostructure MoS$_2$/SiO$_2$ $E^{(0)}_g=2.17\pm0.04$ eV \cite{Rigosi2016} and $\eta_e=2.16\pm0.25$ ($m_e/m_0=0.67\pm0.08$ and $m_h/m_0=0.6\pm0.08$ \cite{Eknapakul2014}). By analogy with EHL \cite{Pekh2020, Pekh2021}, we take the experimentally obtained exciton binding energy in the zero-density limit $|E_\text{ex}|=310\pm40$ meV \cite{Rigosi2016} as a unit of energy measurement. We have in these units $E^{(0)}_g=7\pm1.03$. Then we find $\nu_e\gtrsim61\pm34$. Obviously, this estimate for the number of valleys is larger for encapsulated heterostructures because of the lower binding energy of the exciton. [Using parameters of the heterostructure $h$BN/MoS$_2$/$h$BN, we find $\nu_e\gtrsim119\pm39$.]

We see that the obtained lower bound for the number of valleys required for the transition does indeed correspond to the multivalley case. However, in real heterostructures based on TMD, there are only a few valleys. Usually, there are two valleys at the $K$ points of the Brillouin zone. Valleys at the $\Lambda$ points are populated by intense photoexcitation of TMD bilayers \cite{Chernikov2015b}. There are six of them in the middle of the $\Gamma$---$K$ segments. If the spin degeneracy is removed, they should be counted as 3 (we explicitly take into account in our formulas the spin degeneration $g_s=2$). Then we get $\nu_e=8$ or $\nu_e=5$. Splitting of valleys at $K$ points is also possible for alloys such as Mo$_x$W$_{1-x}$S$_2$ with equivalent energy valleys MoS$_2$ and WS$_2$. The maximum number of valleys for TMD is $\nu_e=10$. We also note that most likely we do not reach such high densities of the electron gas as to descend to the minimum of the ground state energy. For the parameters of the heterostructure MoS$_2$/SiO$_2$, we find $n_{e0}\simeq10^{14}$~cm$^{-2}$ for $\nu_e=2$ and $n_{e0}\simeq10^{15}$ cm$^{-2}$ for $\nu_e=10$.

This indicates the absence of the transition in real heterostructures based on monolayers or bilayers of TMD at zero temperature. The ground state energy of the electron gas (in absolute value) in them turns out to be even lower than the obtained above estimates in the multivalley case. This is a consequence of the substantially large band gap, which increases as the number of monolayers in the film decreases to one. A shining example of this behavior is the platinum disulfide mentioned in Sec.~\ref{s1}.

When considering an $e$-$h$  system at a finite temperature, we recall that in a thermodynamically equilibrium situation, the chemical potential of electrons in both the conduction and the valence bands should be the same. This imposes an additional condition on the chemical potential \cite{Andryushin1977}
\begin{equation}\label{Normalization_condition}
\mu=-E^{(0)}_g.
\end{equation}
The condition \eqref{Normalization_condition} with the chemical potential \eqref{mu_general} determine the dependence of the density $n$ on $T$ and $E^{(0)}_g$.

The chemical potential of the electron gas is written by analogy with \eqref{mu_general}
\begin{equation}\label{mu_general_electron_gas}
\begin{split}
&\mu_e(n_e,\,T)=T\ln\left[\exp\left(\frac{2\pi n_e}{\eta_e\nu_eT}\right)-1\right]\\
&-\frac{2\sqrt{2}}{\sqrt{\pi\nu_e}}\sqrt{n_e}+\frac{\partial}{\partial n_e}\left(n_eE^e_\text{corr}\right).
\end{split}
\end{equation}

\begin{figure}[b!]
\begin{center}
\includegraphics[width=0.5\textwidth]{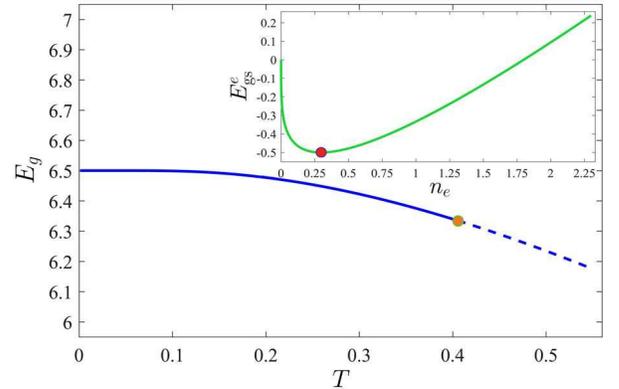}
\caption{\label{f4} Temperature dependence of the band gap in heterostructure MoS$_2$/SiO$_2$ with $\nu_e=2$ at an electron gas density corresponding to the minimum of the ground state energy. The initial band gap is equal to 7. The decrease in the band gap at $T=0$ with respect to 7 is the contribution of the density to the renormalization. The orange point corresponds to the melting temperature of MoS$_2$ (1 458 K). The inset shows the dependence of the ground state energy of the electron gas on density. The red point corresponds to its minimum.}
\end{center}
\end{figure}

To take into account the effect of temperature in the renormalization of the band gap, we should replace $E^e_\text{gs}(n_e)\rightarrow\mu_e(n_e,\,T)$ in \eqref{renorm_band}. We assume now that we have reached at $T=0$ the density $n_{e0}$, and then we increase the temperature. The renormalized band gap is
\begin{equation}\label{renorm_band_T}
E_g=E^{(0)}_g+E^e_0+T\ln\left[1-\exp\left(-\frac{2\pi n_{e0}}{\eta_e\nu_eT}\right)\right],
\end{equation}
where $E^e_0$ is the minimum of $E^e_\text{gs}$.

Fig.~\ref{f4} clearly demonstrates, using heterostructure MoS$_2$/SiO$_2$ as an example, that the contribution of temperature to the renormalization of the band gap turns out to be small compared to the contribution from the electron gas density up to 1 000 K (in dimensionless units, $T\approx0.28$). Taking into account the large value of the dimensionless initial band gap (5--8 for TMD monolayers on the SiO$_2$ substrate and 10--15 for encapsulated heterostructures), it can be confidently asserted that the transition under consideration does not also occur due to an increase in temperature.

\section{\label{s6}Results and conclusions}

We have obtained formulas for the ground state energy of the $e$-$h$ system. We have found the equilibrium density and binding energy of the doped EHL. In the multivalley case, we have derived simple analytical expressions for them. It was shown that the EHL becomes more compressed and strongly coupled upon doping with charge carriers with a larger number of valleys. Estimates have been obtained for the optimal doping level depending on the ratio of the number of electron and hole valleys corresponding to the largest increase in the EHL binding energy. We have also confirmed the band gap renormalization law $\Delta E_g\propto n^{1/3}$ for 2D $e$-$h$ plasmas.

We have considered the gas--liquid transition in heterostructures based on TMD monolayers under electron (hole) doping. The density in the gas phase $n_\text{G}$ and in the liquid phase $n_\text{L}$ form their own bell-shaped curve depending on the doping level $\delta$. The gas--liquid transition at doping corresponds to a surface in the space $(\delta,\,n,\,T)$. The ratios of the equilibrium parameters to the critical ones are weakly dependent on $\delta$; therefore, the critical parameters $T_c$ and $n_c$ depend on $\delta$ approximately in the same way as the binding energy $|E_0|$ and the equilibrium density $n_0$ of EHL on $\delta$.

In the multivalley case, we have obtained analytical expressions for the critical parameters of the gas--liquid transition. We have also obtained temperature dependencies $n_\text{G}(T)$, $n_\text{L}(T)$, and $\mu(T)$ in the vicinity of the critical point and at low temperatures. Formulas are presented with the arbitrary exponent $\gamma$ of the density in the correlation energy ($\gamma=\hspace{0.05cm}^1\hspace{-0.08cm}/_3$ for 2D systems). They are also applicable to layered systems for which $\gamma=\hspace{0.05cm}^1\hspace{-0.08cm}/_4$. In the case of a finite number of valleys, it is also possible to calculate the effective $\gamma$ using the criterion proposed by us earlier.

We have predicted the trion gas--electron-hole plasma transition in heterostructures based on TMD monolayers under electron (hole) doping. This transition takes place in two stages. At the first stage, with increasing density, trions decay into excitons and excess charge carriers. At the second stage, with a further increase in density, excitons decay against the background of a dense electron (hole) gas.

To simplify the consideration of such a transition, we have assumed that initially there is only the trion gas (one excess charge carrier is added to each born $e$-$h$ pair). Firstly, we have estimated the trion gas--electron-hole plasma transition density $n_\text{tr}$ using the Mott criterion, modified for the trion gas. It has been shown that the ratio $n_\text{tr}/n_\text{dm}\approx a^2_B/a^2_\text{tr}$ ($a_B$ and $a_\text{tr}$ are the Bohr radius of 2D exciton and the trion radius, respectively, in the zero density limit). We have obtained by variational calculations for all considered heterostructures based on TMD monolayers $a^2_B/a^2_\text{tr}\simeq\frac{1}{2}$. Then, we have performed variational calculations with a screened Coulomb potential. We have obtained the ratio of densities $n_\text{tr}/n_\text{dm}\approx\frac{2}{3}$. This result refined our estimates of the transition density using the modified Mott criterion and is consistent with the usual metal--insulator transition. We have also get $n_\text{dm}a^2_B\approx0.042-0.044$ and $n_\text{tr}a^2_\text{tr}\approx0.053-0.058$ for all considered heterostructures at $T=0$.

Both transitions occur for the same physical reason. We also recall that in the case of electron (hole) doping, we used the density of the produced $e$-$h$ pairs $n$ as a variable; therefore, the total density of charge carriers together with the excess charge carriers at $\delta=1$ is exactly 1.5 times higher than $n$. Recalculating $n_\text{tr}$ as the total number of particles per unit area, we obtain a value close to $n_\text{dm}$ for the same heterostructure in the absence of doping. This confirms that for screening the Coulomb potential it is not important whether we divide the system of charged particles into $e$-$h$ pairs (excitons) or triples (trions).

We have studied the possibility of the metal--insulator transition in the considered heterostructures in the equilibrium case due to electron (hole) doping. For definiteness, we have assumed that carriers is added to the conduction band. The renormalization of the band gap at $T=0$ is determined by the ground state energy of the electron gas. If the magnitude of the renormalized band gap becomes close to the binding energy of an exciton formed in the presence of the dense electron gas, the band gap abruptly decreases to a certain negative value (the original semiconductor transforms into a semimetal).

However, our calculations have shown that this transition turns out to be impossible even in the multivalley case. Hence, it is impossible in the case of a finite number of valleys at $T=0$. The reason for this is the large initial band gap in TMD monolayers. The contribution to the renormalization of the band gap from temperature is less than from the electron density. It can be stated with confidence that the metal--insulator transition in the equilibrium case is not realized in the considered heterostructures. This means that no this type of the breakdown in devices based on them occurs.

\begin{acknowledgments}
P.~V.~Ratnikov thanks for financial support the Foundation for the Advancement of Theoretical Physics and Mathematics ``BASIS'' (the project no. 20-1-3-68-1).
\end{acknowledgments}


\begin{thebibliography}{60}
\bibitem{RS2018}
P. V. Ratnikov and A. P. Silin, Phys. Usp. {\bf61}, 1139 (2018).
\bibitem{Wang2014}
J. T.-W. Wang \emph{et al.}, Nano Lett. {\bf14}, 724 (2014).
\bibitem{Ahmed2017}
S. Ahmed and J. Yi, Nano-Micro Lett. {\bf9}, 50 (2017).
\bibitem{Dickinson1923}
R. G. Dickinson and L. Pauling, J. Am. Chem. Soc. {\bf45}, 1466 (1923).
\bibitem{Tran2014}
V. Tran, R. Soklaski, Y. Liang, and L. Yang, Phys. Rev. B {\bf89}, 235319 (2014).
\bibitem{Chern2018}
L. A. Chernozatonskii and A. A. Artyukh, Phys. Usp. {\bf61}, 2 (2018).
\bibitem{JeffrisKeldysh1988}
C. D. Jeffries and L. V. Keldysh, eds., \emph{Electron--Hole Droplets in Semiconductors} (Elsevier Science, Amsterdam, 1983).
\bibitem{Yu2019}
Y. Yu \emph{et al.}, ACS Nano {\bf13}, 10351 (2019).
\bibitem{Pekh2020}
P. L. Pekh, P. V. Ratnikov, and A. P. Silin, JETP Lett. {\bf111}, 90 (2020).
\bibitem{Novoselov2004}
K. S. Novoselov, A. K. Geim, S. V. Morozov, D. Jiang, Y. Zhang, S. V. Dubonos, I. V. Grigorieva, and A. A. Firsov, Science {\bf306}, 666 (2004).
\bibitem{Liu2017}
F. Liu, J. Zhou, C. Zhu, and Z. Liu, Adv. Funct. Mater. {\bf27}, 1602404 (2017).
\bibitem{Yu2015}
Y. Yu, F. Yang, X. F. Lu, Y. J. Yan, Y.-H. Cho, L. Ma, X. Niu, S. Kim, Y.-W. Son, D. Feng, S. Li, S.-W. Cheong, X. H. Chen, and Y. Zhang, Nat. Nanotechnol. {\bf10}, 270
(2015).
\bibitem{Kim2017}
B. S. Kim, W. S. Kyung, J. J. Seo, J. Y. Kwon, J. D. Denlinger, C. Kim, and S. R. Park, Sci. Rep. {\bf7}, 5206 (2017).
\bibitem{Rama2011}
A. Ramasubramaniam, D. Naveh, and E. Towe, Phys. Rev. B {\bf84}, 205325 (2011).
\bibitem{Chu2015}
T. Chu, H. Ilatikhameneh, G. Klimeck, R. Rahman, and Z. Chen, Nano Lett. {\bf15}, 8000 (2015).
\bibitem{Mak2013}
K. F. Mak, K. He, C. Lee, G. H. Lee, J. Hone, T. F. Heinz, and J. Shan, Nat. Mater. {\bf12}, 207 (2013).
\bibitem{Ross2013}
J. S. Ross, S. Wu, H. Yu, N. J. Ghimire, A. M. Jones, G. Aivazian, J. Yan, D. G. Mandrus, D. Xiao, W. Yao, and X. Xu, Nat. Commun. {\bf4}, 1474 (2013).
\bibitem{Yang2015}
J. Yang, T. L\"{u}, Y. W. Myint, J. Pei, D. Macdonald, J.-C. Zheng, and Y. Lu, ACS Nano {\bf9}, 6603 (2015).
\bibitem{Chernikov2015a}
A. Chernikov, A. M. van der Zande, H. M. Hill, A. F. Rigosi, A. Velauthapillai, J. Hone, and T. F. Heinz, Phys. Rev. Lett. {\bf115}, 126802 (2015).
\bibitem{Pekh2021}
P. L. Pekh, P. V. Ratnikov, and A. P. Silin, JETP {\bf133}, 494 (2021).
\bibitem{Andryushin1976a}
E. A. Andryushin, V. S. Babichenko, L. V. Keldysh, T. A. Onishchenko, and A. P. Silin, JETP Lett. {\bf24}, 185 (1976).
\bibitem{Andryushin1990a}
E. A. Andryushin and A. P. Silin, Sov. Phys. Solid State {\bf32}, 1746 (1990).
\bibitem{Andryushin1990b}
E. A. Andryushin and A. P. Silin, Fiz. Tverd. Tela {\bf32}, 3579 (1990).
\bibitem{Andryushin1991}
E. A. Andryushin and A. P. Silin, J. Moscow Phys. Soc. {\bf1}, 59 (1991).
\bibitem{Durnev2018}
M. V. Durnev and M. M. Glazov, Phys. Usp. {\bf61}, 825 (2018).
\bibitem{Andryushin1976b}
E. A. Andryushin and A. P. Silin, Sov. Phys. Solid State {\bf18}, 1243 (1976).
\bibitem{Andryushin1976c}
E. A. Andryushin and A. P. Silin, Solid State Comm. {\bf20}, 453 (1976).
\bibitem{Tikhodeev1985}
S. G. Tikhodeev, Sov. Phys. Usp. {\bf28}, 1 (1985).
\bibitem{Trankle1987}
G. Tr\"{a}nkle, E. Lach, A. Forchel, C. Ell, H. Haug, G. W. G. Griffiths, H. Kroemer, and S. Subbanna, J. Phys. Colloq. {\bf48}, 385 (1987).
\bibitem{Andryushin1979}
E. A. Andryushin and A. P. Silin, Sov. Phys. Solid State {\bf21}, 129 (1979).
\bibitem{Andryushin1977}
E. A. Andryushin, L. V. Keldysh, and A. P. Silin, Sov. Phys. JETP {\bf46}, 616 (1977).
\bibitem{Kraeft1986}
W.-D. Kraeft, D. Kremp, W. Ebeling, and G. R\"{o}pke, \emph{Quantum Statistics of Charged Particle Systems} (Akademie-Verlag, Berlin, 1986).
\bibitem{Mott1968}
N. F. Mott, Rev. Mod. Phys. {\bf40}, 677 (1968).
\bibitem{Mott1972}
N. F. Mott, Adv. Phys. {\bf21}, 785 (1972).
\bibitem{Mott1973}
N. F. Mott, Contemp. Phys. {\bf14}, 401 (1973).
\bibitem{Edwards1978}
P. P. Edwards and M. J. Sienko, Phys. Rev. B {\bf17}, 2575 (1978).
\bibitem{Courtade2017}
E. Courtade, M. Semina, M. Manca, M. M. Glazov, C. Robert, F. Cadiz, G. Wang, T. Taniguchi, K. Watanabe, M. Pierre, W. Escoffier, E. L. Ivchenko, P. Renucci,
X. Marie, T. Amand, and B. Urbaszek, Phys. Rev. B {\bf96}, 085302 (2017).
\bibitem{Glazov2018}
M. M. Glazov and A. Chernikov, Phys. Stat. Sol. B {\bf255}, 1800216 (2018).
\bibitem{Rytova1967}
N. S. Rytova, Moscow Univ. Phys. Bulletin {\bf22}, 18 (1967).
\bibitem{Chaplik1971}
A. V. Chaplik and M. V. Entin, Sov. Phys. JETP {\bf34}, 1335 (1971).
\bibitem{Keldysh1979}
L. V. Keldysh, JETP Lett. {\bf29}, 658 (1979).
\bibitem{Chernikov2015b}
A. Chernikov, C. Ruppert, H. M. Hill, A. F. Rigosi, and T. F. Heinz, Nat. Photon. {\bf9}, 466 (2015).
\bibitem{Eknapakul2014}
T. Eknapakul, P. D. C. King, M. Asakawa, P. Buaphet, R.-H. He, S.-K. Mo, H. Takagi, K. M. Shen, F. Baumberger, T. Sasagawa, S. Jungthawan, and W. Meevasana,
Nano Lett. {\bf14}, 1312 (2014).
\bibitem{Berkelbach2013}
T. C. Berkelbach, M. S. Hybertsen, and D. R. Reichman, Phys. Rev. B {\bf88}, 045318 (2013).
\bibitem{Rasmussen2015}
F. A. Rasmussen and K. S. Thygesen, J. Phys. Chem. C {\bf119}, 13169 (2015).
\bibitem{Goryca2019}
M. Goryca, J. Li, A. V. Stier, T. Taniguchi, K. Watanabe, E. Courtade, S. Shree, C. Robert, B. Urbaszek, X. Marie, and S. A. Crooker, Nat. Commun. {\bf10}, 4172 (2019).
\bibitem{Zipfel2020}
J. Zipfel, M. Kulig, R. Perea-Caus´ın, S. Brem, J. D. Ziegler, R. Rosati, T. Taniguchi, K. Watanabe, M. M. Glazov, E. Malic, and A. Chernikov, Phys. Rev. B {\bf101},
115430 (2020).
\bibitem{Andryushin1981}
E. A. Andryushin, A. P. Silin, and V. A. Sanina, Fiz. Tverd. Tela {\bf23}, 1200 (1981).
\bibitem{Silin1988}
A. P. Silin, Trudy FIAN {\bf118}, 113 (1988).
\bibitem{Hubbard1957}
J. Hubbard, Proc. Roy. Soc. {\bf A243}, 336 (1957).
\bibitem{Silin1978}
A. P. Silin, Fiz. Tverd. Tela {\bf20}, 3436 (1978).
\bibitem{Rigosi2016}
A. F. Rigosi, H. M. Hill, K. T. Rim, G. W. Flynn, and T. F. Heinz, Phys. Rev. B {\bf94}, 075440 (2016).
\end{thebibliography}
\end{document}